%% file: SUS-14-010_temp.tex
\pdfoutput=1

\documentclass[11pt,twoside,a4paper,cmspaper,final,collab]{cms-tdr}
\def\svnVersion{285915}\def\cmsCernNo{2013-037}\def\cmsCernDate{\today}\def\cmsMessage{Published in Physics Letters B as \href{http://dx.doi.org/10.1016/j.physletb.2015.04.002}{\doi{10.1016/j.physletb.2015.04.002.}}}

\begin{document}\cmsNoteHeader{SUS-14-010}

\hyphenation{had-ron-i-za-tion}
\hyphenation{cal-or-i-me-ter}
\hyphenation{de-vices}
\RCS$Revision: 283795 $
\RCS$HeadURL: svn+ssh://svn.cern.ch/reps/tdr2/papers/SUS-14-010/trunk/SUS-14-010.tex $
\RCS$Id: SUS-14-010.tex 283795 2015-04-08 19:56:56Z alverson $
\newlength\cmsFigWidth
\ifthenelse{\boolean{cms@external}}{\setlength\cmsFigWidth{0.98\columnwidth}}{\setlength\cmsFigWidth{0.6\textwidth}}
\ifthenelse{\boolean{cms@external}}{\providecommand{\cmsLeft}{top\xspace}}{\providecommand{\cmsLeft}{left\xspace}}
\ifthenelse{\boolean{cms@external}}{\providecommand{\cmsRight}{bottom\xspace}}{\providecommand{\cmsRight}{right\xspace}}

\newcommand{\stlep}{\ensuremath{S_{\cmsSymbolFace{T}}^{\text{lep}}}\xspace }
\newcommand{\njets}{\ensuremath{N_{\text{jets}}}\xspace}
\newcommand{\mht}{\ensuremath{H_{\text{T}}^{\text{miss}}}\xspace}

\newcommand{\nbjets}{\ensuremath{N_{\cPqb \text{jets}}}\xspace}
\newcommand{\bjets}{\ensuremath{\cPqb \text{ jets}}\xspace}
\cmsNoteHeader{SUS-14-010}
\title{Searches for supersymmetry based on events with \bjets and four \PW\ bosons in pp collisions at 8 TeV}

\date{\today}

\abstract{
Five mutually exclusive searches for supersymmetry are presented based on events in which \bjets and four \PW\ bosons are produced in proton-proton collisions at $\sqrt{s}=8\TeV$. The data, corresponding to an integrated luminosity of 19.5\fbinv, were collected with the CMS experiment at the CERN LHC in 2012.  The five studies differ in the leptonic signature from the W boson decays, and correspond to all-hadronic, single-lepton, opposite-sign dilepton, same-sign dilepton, and $\geq$3 lepton final states. The results of the five studies are combined to yield 95\% confidence level limits for the gluino and bottom-squark masses in the context of gluino and bottom-squark pair production, respectively. In the limit when the lightest supersymmetric particle is light, gluino and bottom squark masses are excluded below 1280 and 570\GeV, respectively.
}

\hypersetup{%
pdfauthor={CMS Collaboration},%
pdftitle={Searches for supersymmetry based on events with b jets and four W bosons in pp collisions at 8 TeV},%
pdfsubject={CMS},%
pdfkeywords={CMS, physics, supersymmetry}}

\maketitle
\section{Introduction}
The standard model  (SM) of particle physics provides an accurate description of known particle properties and interactions. The discovery of a Higgs boson by the ATLAS \cite{Aad:2012tfa}  and CMS~\cite{Chatrchyan:2012ufa} Collaborations at the CERN LHC represents the latest
  major milestone in the validation of the SM.  Despite its success, the SM is known to be incomplete because, for example, it does not offer an explanation for dark matter and it contains ad-hoc  features, such as the fine-tuning~\cite{Barbieri:1987fn,Romanino:1999ut,Feng:1999mn,Kitano:2005wc,Giudice:2006sn,Barbieri:2009ev,Horton:2009ed} required to stabilize the Higgs boson mass at the electroweak scale. Many extensions to the SM have been proposed. In particular, supersymmetry (SUSY) may provide a candidate for dark matter in $R$-parity conserving models~\cite{Farrar:1978xj} as well as a natural solution to the fine-tuning problem~\cite{Barbieri:1987fn,Romanino:1999ut,Feng:1999mn,Kitano:2005wc,Giudice:2006sn,Barbieri:2009ev,Horton:2009ed}.

The CMS and ATLAS Collaborations have performed many searches for physics beyond the SM. Thus far, no significant evidence for new physics has been obtained. The search for supersymmetry is particularly interesting phenomenologically because of the
large number of new particles expected. The LHC SUSY search program consists therefore of a wide array of searches~\cite{Aad:2013wta,Aad:2013ija,Aad:2014nua,Aad:2014qaa,Aad:2014vma,Aad:2014iza,Chatrchyan:2013wxa,Chatrchyan:2013xna,Chatrchyan:2014lfa,Chatrchyan:2013iqa,Chatrchyan:2013fea,Khachatryan:2014doa}. Any particular
manifestation of SUSY in nature would likely result in topologies that are detectable in a variety of final-states. Individual searches can therefore be combined to provide complementarity and enhanced sensitivity in the global search
for new physics.

Naturalness arguments suggest that the supersymmetric partners of the gluon (gluino, $\PSg$) and third-generation quarks (the top and bottom squarks, \sTop and \sBot) should not be too heavy~\cite{Sakai:1981gr,Dimopoulos:1995mi,Papucci:2011wy,Brust:2011tb}. Direct or cascade production of third-generation squarks can lead to final states with several \PW\ bosons and bottom quarks, and considerable imbalance \ptvecmiss in transverse momentum.  The missing momentum arises from neutrinos in events where one or more W bosons decay leptonically, but also, for the R-parity conserving models considered here, because the lightest SUSY particle (LSP), taken to be the lightest neutralino $\PSGczDo$, is weakly interacting and stable,  escaping without detection.  The studies presented here focus on  SUSY simplified model scenarios~\cite{sms1,sms2}  with four \PW\ bosons.  Each of the \PW\ bosons can decay either into a quark-antiquark pair or into a charged lepton and its neutrino. Depending on the decay modes of the \PW\ bosons, the final states contain 0--4 leptons. This makes combining the final states with different lepton multiplicities beneficial. The dilepton signature is split according to the relative electric charges of the leptons, providing five mutually exclusive analyses for the combination: fully hadronic, single-lepton, opposite-sign dilepton, same-sign dilepton, and $\geq$3 leptons (multilepton). The results are based on proton-proton collision data collected at $\sqrt{s}=8$\TeV with the CMS experiment at the LHC during 2012, and correspond to an integrated luminosity of 19.5\fbinv.

\begin{figure*}[bthp]
\centering
\includegraphics[width=0.32\textwidth]{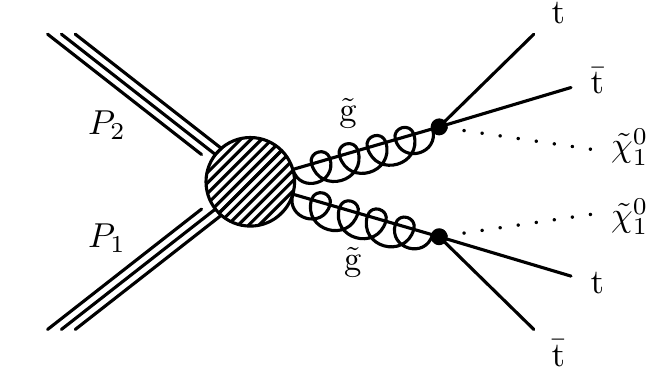}
\includegraphics[width=0.32\textwidth]{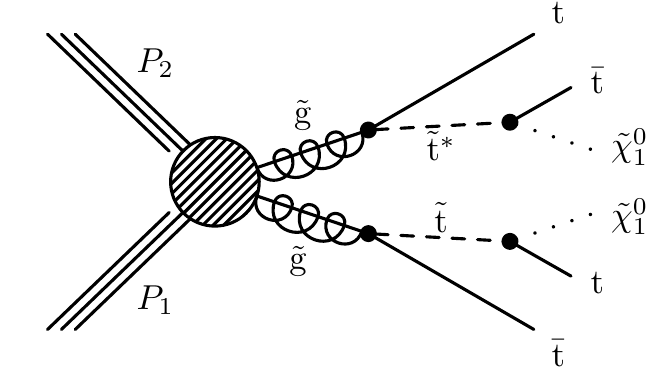}
\includegraphics[width=0.32\textwidth]{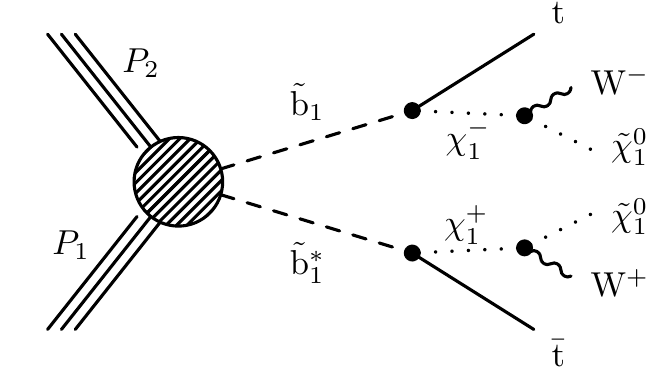}
\caption{Feynman diagrams for the signals from: (left)~gluino pair production with intermediate virtual top squarks (T1tttt), (middle)~gluino pair production with intermediate on-shell top squarks (T5tttt), and (right)~bottom squark pair production (T6ttWW).
\label{fig:feyn}}
\end{figure*}

The first simplified model we consider describes gluino pair production, followed by the decay of each gluino to a top quark-antiquark pair (\ttbar) and the LSP. For cases where the top squark mass is larger than the gluino mass, the decay will proceed through a virtual top squark (T1tttt model, Fig.~\ref{fig:feyn}\,left). Alternatively, when the top squark mass is smaller than the gluino mass and phase space allows, the decay will proceed through an on-shell top squark (T5tttt model, Fig.~\ref{fig:feyn}\,middle). Each top quark decays to a bottom quark and a \PW\ boson, leading to final states with four W bosons, four bottom-quark jets (\bjets), and considerable \ptvecmiss. The second simplified model we consider describes bottom-antibottom squark pair production, where we assume that each bottom squark decays to a top quark and a chargino (\chipm), and that the chargino then decays to yield a W boson and the LSP (T6ttWW model, Fig.~\ref{fig:feyn}\,right). The final state thus contains four W bosons, two \bjets, and large \ptvecmiss.

The paper is organized as follows. Section 2 describes the CMS detector.  The event simulation, trigger, and reconstruction procedures are described in Section 3.
Section 4 presents details of the individual analyses, with particular emphasis on the opposite-sign dilepton search, which is presented here for the first time. The
combination methodology and results are presented in Section 5. Section 6 provides a summary.

\section{Detector}

The central feature
of the CMS detector is a superconducting solenoid, of 6\unit{m} internal
diameter, that produces an axial magnetic field of 3.8\unit{T}.  Within the
solenoid volume are a silicon pixel and strip tracker, a lead tungstate
crystal electromagnetic calorimeter, and a brass and plastic
scintillator hadron calorimeter. Muons are detected in gas ionization chambers embedded in the steel flux-return yoke outside the magnet. The tracking system covers the pseudorapidity range $\abs{\eta}<2.5$, the muon detectors $\abs{\eta}<2.4$, and the calorimeters $\abs{\eta}<3.0$. Steel and quartz-fiber forward calorimeters cover $3<\abs{\eta}<5$. A detailed description of the CMS apparatus and coordinate  system are given in Ref.~\cite{CMS}.

\section{Event reconstruction, trigger, and simulation}

The recorded events are reconstructed using the CMS particle-flow algorithm~\cite{PFT-09-001,PFT-10-001}.
Electron candidates are reconstructed by associating tracks
to energy clusters in the electromagnetic calorimeter~\cite{EGM-10-004,Chatrchyan:2013dga}. Muon candidates are reconstructed by
combining information from the tracker and the muon detectors~\cite{CMS-PAPER-MUO-10-004}.

Particle-flow constituents are clustered into jets using the anti-\kt clustering algorithm
with a distance parameter of 0.5~\cite{antikt}. Corrections are applied as a function of jet transverse momentum (\pt) and $\eta$ to account for non-uniform detector response~\cite{JETJINST,METPAS}. Contributions from additional pp collisions overlapping with the event of interest (pileup) are estimated using the jet area method~\cite{PU_JET_AREAS,JET_AREAS} and are subtracted from the jet \pt. The total visible jet activity \HT is defined as the scalar sum of the jet \pt in the event, and \mht as the \pt imbalance of the reconstructed jets, where the \pt and $\eta$ requirements for accepted jets are specified for the individual searches in Section 4.
The identification of \bjets is performed using the combined secondary vertex algorithm at the medium working point~\cite{ref:btag}, which has a b-jet tagging efficiency of 70\% and a light-flavor jet misidentification rate below 2\% for jets with \pt values in the range of interest for this analysis.  The missing transverse momentum vector \ptvecmiss is  defined as the projection on the plane perpendicular to the beam axis of the negative vector sum of the momenta of all reconstructed  particles. Its magnitude is referred to as \ETmiss.

The data sample used for the fully hadronic analysis was recorded with trigger algorithms that required events to have $\HT> 350$\GeV and $\ETmiss > 100$\GeV.  The single-lepton analysis uses triple- or double-object triggers.
The triple-object triggers require a lepton with $\pt>15$\GeV, together with $\HT>350$\GeV and $\ETmiss > 45$\GeV. The double-object triggers have the same \HT requirement, no $\ETmiss$ requirement, and a lepton \pt threshold of 40\GeV.  The data samples for the dilepton and multilepton analyses were collected with $\Pe\Pe$, $\Pe\Pgm$, and $\Pgm\Pgm$ double-lepton triggers, which require at least one $\Pe$ or one $\Pgm$ with $\pt>17\GeV$ and another with $\pt>8\GeV$.

Simulated Monte Carlo (MC) samples of signal events are produced  using the \MADGRAPH 5.1.3.30~\cite{ref:ref_019}  generator, as are SM \ttbar, Drell--Yan, \PW+jets, and single top quark events.  The \ttbar events include production in association with a photon, or with a \PW, \cPZ, or \PH\ boson. The production of single top quarks in association with an additional quark and a \cPZ\ boson is simulated with the \MCATNLO 2.0.0~\cite{MCatNLO1,MCatNLO2} generator. The PYTHIA 6.4.24~\cite{Sjostrand:2006za} generator is used to simulate the generic multijet QCD and diboson (\PW\PW, \cPZ\cPZ, and \PW\cPZ) processes,  as well as to describe the parton shower and hadronization for the \MADGRAPH samples. All SM samples are processed with the full simulation of the CMS detector,
based on the \GEANTfour~\cite{Geant} package, while the signal samples are processed with the CMS fast simulation~\cite{Abdullin:2011zz} program. The fast simulation is
validated through comparison of its predictions with those of the full simulation, and efficiency corrections based on data are applied~\cite{Rahmat:2012fs}.  The effect of pileup
interactions is included by superimposing a number of simulated minimum bias events on top of the
hard-scattering process, with the distribution of the number of reconstructed vertices matching that in data.
\section{Search channels}
 This paper reports the combination of five individual searches for new physics by CMS. The fully hadronic~\cite{Chatrchyan:2014lfa}, single-lepton~\cite{Chatrchyan:2013iqa}, same-sign
 dilepton~\cite{Chatrchyan:2013fea}, and multilepton~\cite{Khachatryan:2014doa} searches have all been
 published, and are summarized briefly below. The opposite-sign dilepton search is presented  here for the first time and is therefore described in greater detail.

\subsection{Fully hadronic analysis}

Considering that signal events contain four \PW\ bosons, the fully hadronic branching fraction is about 24\%. The fully hadronic analysis~\cite{Chatrchyan:2014lfa} requires at least three jets with $\pt > 50$\GeV and $\abs{\eta}<2.5$, and vetoes events containing an isolated electron or muon with $\pt>10\GeV$ and $\abs{\eta}<2.5$\,(2.4) for electrons (muons). The \HT and \mht values are required to exceed 500 and 200\GeV, respectively.  To render the analysis more sensitive to a variety of final-state topologies resulting from longer cascades of squarks and gluinos, and therefore
a large number of jets, the events are divided into
three exclusive jet-multiplicity regions: $\njets = ($3--5), (6--7), and $\geq$8.  The events are further divided into exclusive regions of  \HT and \mht.  The exploitation of  higher jet multiplicities
is motivated by natural SUSY models in which the gluino decays into top quarks~\cite{Chatrchyan:2014lfa}. This analysis does not impose a requirement on the number \nbjets of tagged \bjets, thereby maintaining a high signal efficiency.

The main SM backgrounds for the fully hadronic channel arise from \Z+jets events in which the \Z\ boson decays to a $\Pgn\Pagn$ neutrino pair; from \PW+jets and \ttbar events with a \PW\ boson that decays directly or through a $\tau$ lepton to an e or $\Pgm$ and the associated neutrino(s), with the e or $\Pgm$ undetected or outside the acceptance of the analysis; from \PW+jets and \ttbar events with a \PW\ boson that decays to a hadronically decaying $\tau$ lepton and its associated neutrino; and from QCD multijet events. For the first three background categories, the neutrinos provide a source of genuine \mht.  For the QCD multijet event background, large values of \mht arise from the mismeasurement of jet \pt or from the neutrinos produced in the semileptonic decays of hadrons. All SM backgrounds are determined from control regions in the data,  and are found to agree with the observed numbers of events in the signal regions.

\subsection{Single-lepton analysis}

With four \PW\ bosons, the branching fraction of signal events to states with a single electron or muon is about 42\%, including contributions from leptonically decaying $\tau$ leptons.  The single-lepton analysis~\cite{Chatrchyan:2013iqa} requires the presence of  an electron or muon with $\pt>20\GeV$ and no second electron or muon with $\pt>15\GeV$, with the same $\eta$  restrictions on the e and $\Pgm$ as in Section 4.1.  Jets are required to have $\pt>40\GeV$ and $\abs{\eta}<2.4$. The $S_{\mathrm{T}}^{\text{lep}}$ variable is evaluated, defined by the scalar sum of \ETmiss and the lepton \pt. Events must satisfy $\njets \geq 6$, $\nbjets \geq 2$, $\HT>400\GeV$, and $S_{\mathrm{T}}^{\text{lep}}>250\GeV$. A further variable, the azimuthal angle $\Delta\phi (\PW,\ell)$ between the \PW\ boson candidate and the lepton, is evaluated.  For this variable,  the \pt of the W boson candidate is defined by the  vector sum of the lepton \pt and $\ptvecmiss$. For single-lepton \ttbar events, the angle between the directions of the \PW\ boson and the charged lepton has a maximum value that is determined by the mass of the \PW\ boson and its momentum. The requirement of large \ETmiss selects events with Lorentz-boosted \PW\ bosons.  This leads to a narrow distribution in $\Delta\phi (\PW,\ell)$.  In SUSY decays there will be no such maximum, since the \ETmiss mostly results from the two neutralinos and their directions are largely independent of the lepton direction.  Therefore the $\Delta\phi (\PW,\ell)$ distribution is expected to be flat for SUSY events. The analysis requires $\Delta\phi (\PW,\ell) > 1$.  The search is then performed in exclusive regions of $S_{\mathrm{T}}^{\text{lep}}$ for $\nbjets=2$ and $\nbjets\geq3$.

The main SM backgrounds for the single-lepton channel arise from dilepton \ttbar events in which one lepton is not reconstructed or lies outside the acceptance of the analysis, from residual single-lepton \ttbar events, and from events with single-top quark production.  The backgrounds are evaluated using data control samples.  The total  number of background events is found to agree with the observed  number of events in each signal region.

\subsection{Same-sign dilepton analysis} 

The branching fraction for events with four \PW\ bosons to a final state with at least two same-sign leptons ($\Pe\Pe$, $\Pgm\Pgm$, or $\Pe\Pgm$) is 7\%, including the contributions of $\tau$ leptons. For the present study, we make use of the high-\pt selection of the same-sign dilepton analysis in Ref.~\cite{Chatrchyan:2013fea}, which
  requires at least two same-sign light leptons ($\Pe$, $\Pgm$) with $\pt>20\GeV$,  $\abs{\eta}<2.4$, and invariant mass above 8\GeV.
To prevent overlap between the same-sign dilepton and multilepton analyses, an explicit veto on additional leptons with $\pt>10\GeV$ and  $\abs{\eta}<2.4$ is added for the same-sign dilepton analysis, as in the search for
$\PSQt_{2}$ production described in Ref.~\cite{Khachatryan:2014doa}.
Jets are required to satisfy $\pt>40\GeV$ and $\abs{\eta}<2.4$. Events must have $\njets>2$, $\HT>200\GeV$, and $\ETmiss>50\GeV$. The events are examined in exclusive regions of \HT and \ETmiss for $2\leq \njets\leq 3$ and $\njets\geq 4$, all for $\nbjets = 0$, 1, and $\geq$2.

There are three main sources of SM background in this analysis: non-prompt leptons, rare SM processes, and electrons with wrong charge assignments.
The main sources of non-prompt leptons are leptons from bottom- and charm-quark decays, misidentified hadrons, muons from light-meson decays in flight, and electrons
from unidentified photon conversions.  The background from non-prompt leptons is evaluated from data control regions. Diboson, $\ttbar\PW$,  and $\ttbar\cPZ$ production are the most important rare SM background sources.  Their contributions are estimated from MC simulation. Opposite-sign dileptons can also contribute to the background when the charge of an electron is misidentified because of bremsstrahlung emitted in the tracker material. This contribution is estimated using a technique based on $Z\to\,\text{e}^+\text{e}^-$ data. No significant deviations are observed from the SM expectations.

\subsection{Multilepton analysis} 

The branching fraction for events with four W bosons to decay to a final state with three or more charged leptons ($\Pe$ or $\Pgm$)  is 6\%, including $\tau$ lepton contributions. The multilepton sample used in the present study corresponds to the selection of events with three or more such leptons presented in Ref.~\cite{Khachatryan:2014doa}.   The electrons or muons are required to have $\pt>10\GeV$ and $\abs{\eta}<2.4$, except at least one of the three leptons must have
  $\pt>20\GeV$. Jets are required to have $\pt>30\GeV$ and $\abs{\eta}<2.4$. Events must satisfy $\njets\geq 2$, $\nbjets\geq 1$, $\HT>60\gev$, and $\ETmiss>50\GeV$.
  The events are examined in exclusive regions of \HT and \ETmiss for $2\leq\njets\leq 3$ and $\njets\geq 4$, both with $\nbjets = 1$ and 2, and for $\njets \geq 3$ with $\nbjets\geq 3$.

Compared to the fully hadronic, single-lepton, or dilepton signatures, the multilepton search targets final states with
small branching fractions, but provides good signal sensitivity because the three-lepton requirement strongly
suppresses backgrounds. Only a few SM processes exhibit such signatures. Background from diboson production is highly suppressed by the \nbjets requirement.  The main backgrounds arise from events with a combination of \ttbar production and non-prompt leptons, as well as from rare SM processes like {\ttbar}W and {\ttbar}Z production. The non-prompt lepton background is evaluated using data control  regions and the rare SM background from simulation.  There is no  statistically significant excess of events found in the signal regions above the SM expectations.

\subsection{Opposite-sign dilepton analysis}

The branching fraction for events with four \PW\ bosons to a final state with at least one opposite-sign lepton pair ($\Pe$ or $\Pgm$) is 14\%, including the
 contributions of $\tau$ leptons.  The opposite-sign dilepton search requires the presence of exactly two opposite-sign leptons (e or $\Pgm$), each with $\pt>20\GeV$ and $\abs{\eta}<2.4$. Events with a third lepton satisfying $\pt>20\GeV$ and  $\abs{\eta}<2.4$ are vetoed. Jets must satisfy $\pt>30\GeV$ and $\abs{\eta}<2.5$. This analysis targets the T1tttt and T5tttt scenarios described in the Introduction.

 Many variables are examined in order to define a signal region (SR) that maximizes signal content while minimizing the contributions of SM events. We choose those variables that demonstrate the greatest discriminating
  power between signal and SM events, and that  exhibit the smallest level of correlation amongst themselves: \njets, \nbjets, \ETmiss, and the $\eta$ values of the two jets with largest \pt.
 The criteria that yield the highest sensitivity in the parameter space of the T1tttt model, summarized in
 Table \ref{table1}, are optimized using simulated events. Events
 are divided into bins of \ETmiss. The bin with highest \ETmiss ($>$180\GeV) is the most sensitive for the bulk of the signal phase space, but the bins with lower \ETmiss are important  for compressed spectra, \ie, for signal scenarios with small mass differences between the SUSY particles. After applying the selection criteria summarized in Table~\ref{table1}, the remaining SM background is primarily composed of events with \ttbar, Drell--Yan, and \PW+jets production.

\begin{table*}[htb]
\centering
\topcaption{Selection criteria for the signal region in the opposite-sign dilepton analysis.}
\begin{tabular}{ccc}
 Variable & Description & Criterion \\ \hline
 \ETmiss                          & Missing transverse momentum           & $>$30\GeV  \\
 $N_\text{jets}$                   & Number of jets                      & $>$4        \\
 \nbjets                 & Number of \cPqb-tagged jets             & $>$2        \\
 $\abs{\eta_{j_{1}}}$                     & Pseudorapidity for jet with largest \pt              & $<$1          \\
 $\abs{\eta_{j_{2}}}$                     & Pseudorapidity for jet with next-to-largest \pt                  & $<$1          \\
\end{tabular}
\label{table1}
\end{table*}

A control region (CR) is defined by the sum of the two event samples obtained by separately inverting the $\eta_{j_{1}}<1$ and $\eta_{j_{2}}<1$ requirements.  The contribution of signal events to the control region depends on the gluino mass ($m_{\PSg}$) and the LSP mass ($m_\text{LSP}$) and can be as large as 10\%. The contributions of signal events to the CR are taken into account in the interpretation of the results.

\begin{figure}[htbp]
   \centering
      \includegraphics[width=0.5\textwidth]{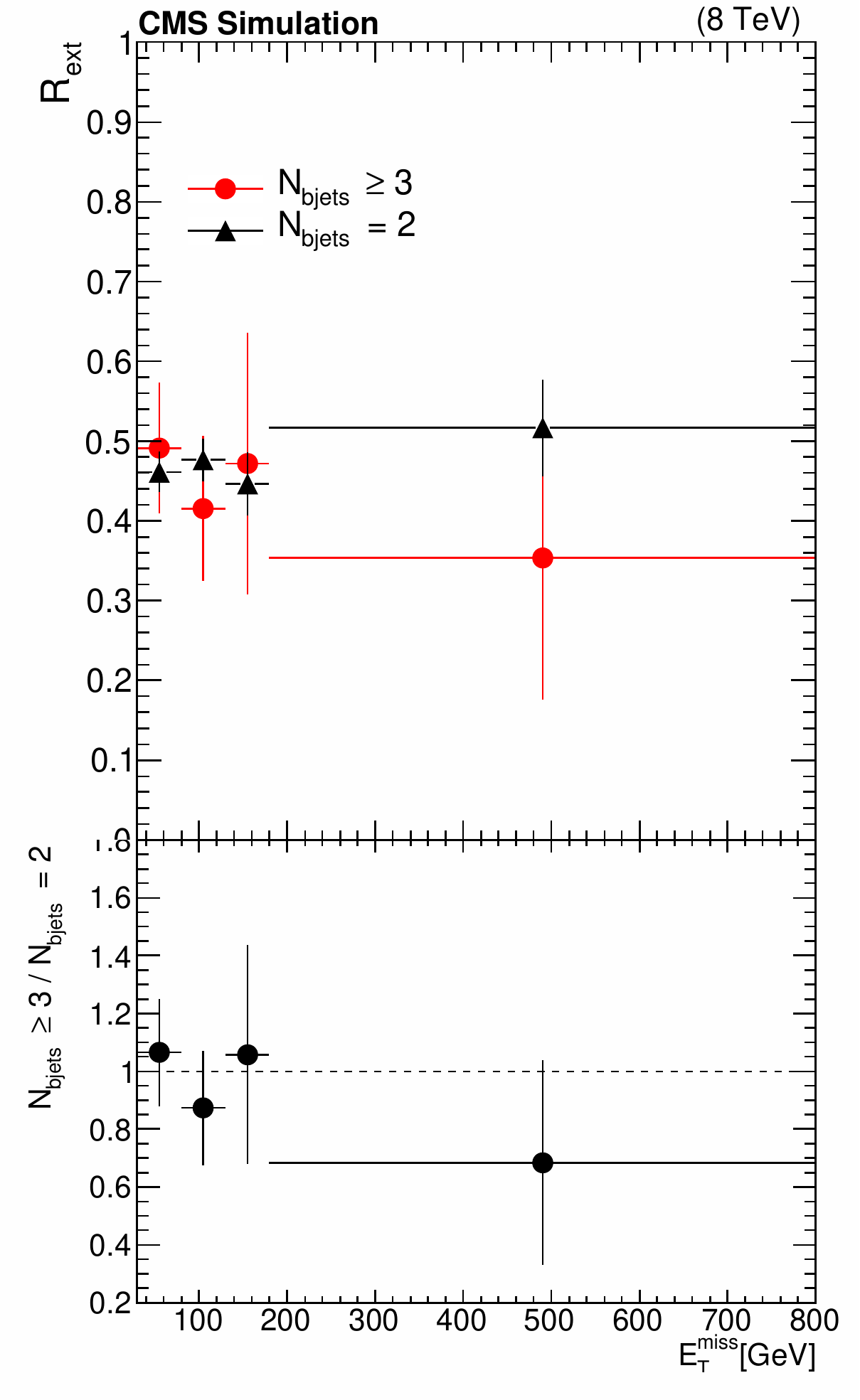}
      \caption{Extrapolation factors from the control region to the signal region, $R_\text{ext}$, as a function of \ETmiss, for simulated events with $\nbjets=2$ (black triangles) and $\nbjets\geq3$ (red points).  All the other signal selection criteria have been applied. The lower panel shows the ratio of the $\nbjets\geq3$ to the $\nbjets=2$ results. }
      \label{fig2}
 \end{figure}

An extrapolation factor $R_\text{ext}$ is defined as a function of \ETmiss and $\nbjets$, as the ratio of the number of SM events in the SR  to that in the CR.
In simulated events the $R_\text{ext}$ factor is observed to change slowly as a function of \ETmiss, as shown in Fig.~\ref{fig2}. The $R_\text{ext}$ ratio is similarly found to be independent of \nbjets,  making it possible to extract its value directly from data using events with $\nbjets=2$, without altering the other signal and control selection criteria.  The contribution of signal events to the $\nbjets=2$ region is small compared to the statistical uncertainty in the extrapolation factor and is therefore neglected.
Thus the background estimate is derived entirely from data, minimizing systematic uncertainties.

The SM background prediction for the SR is obtained by multiplying $R_\text{ext}$ with the number of data events in the CR:
\begin{equation}
 N_{\text{predicted}}^{\text{SR}} = R_\text{ext} N_{\text{data}}^{\text{CR}}.
\end{equation}

\begin{figure*}[htbp]
   \centering
       \includegraphics[width=0.32\textwidth]{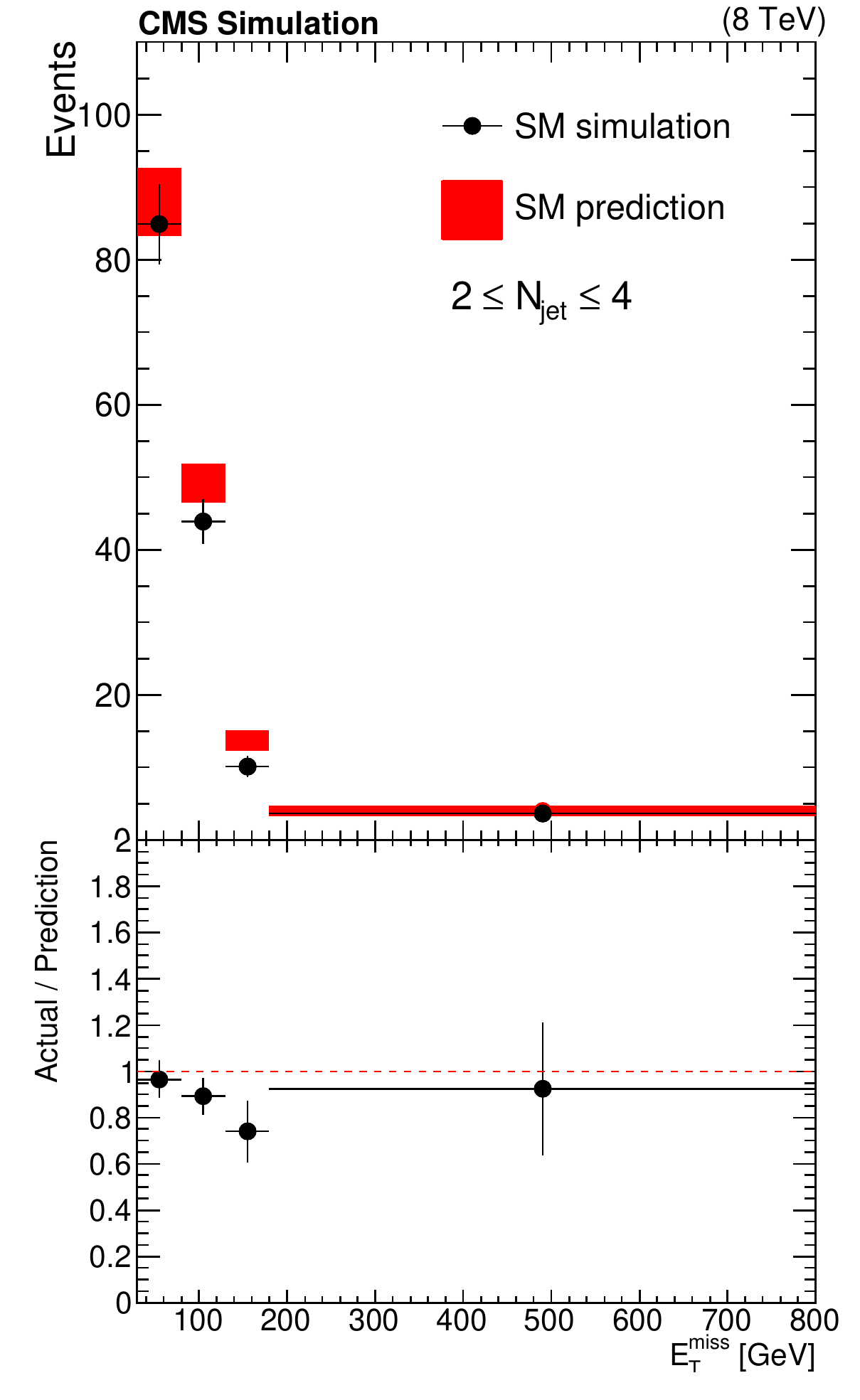}
       \includegraphics[width=0.32\textwidth]{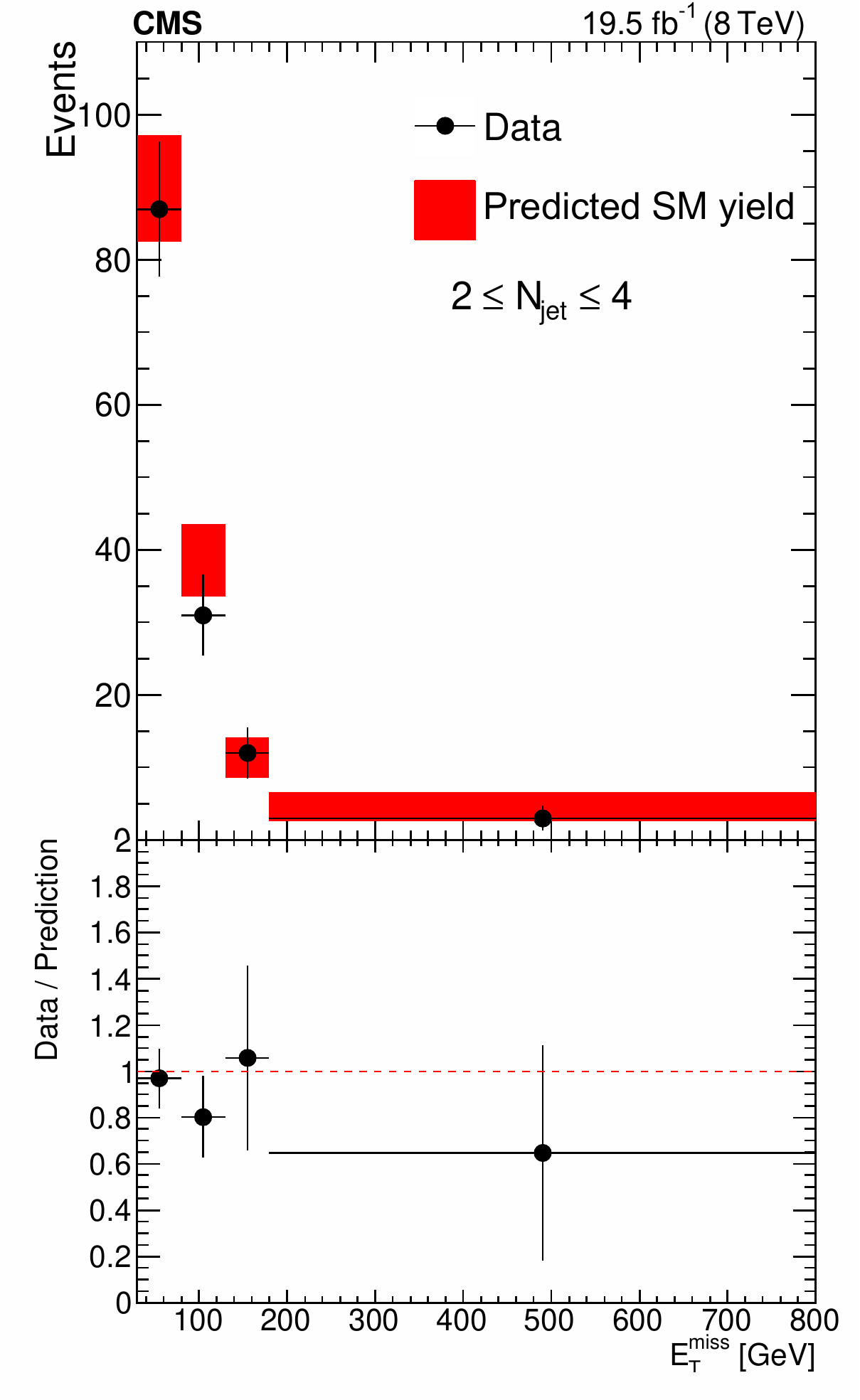}
       \includegraphics[width=0.32\textwidth]{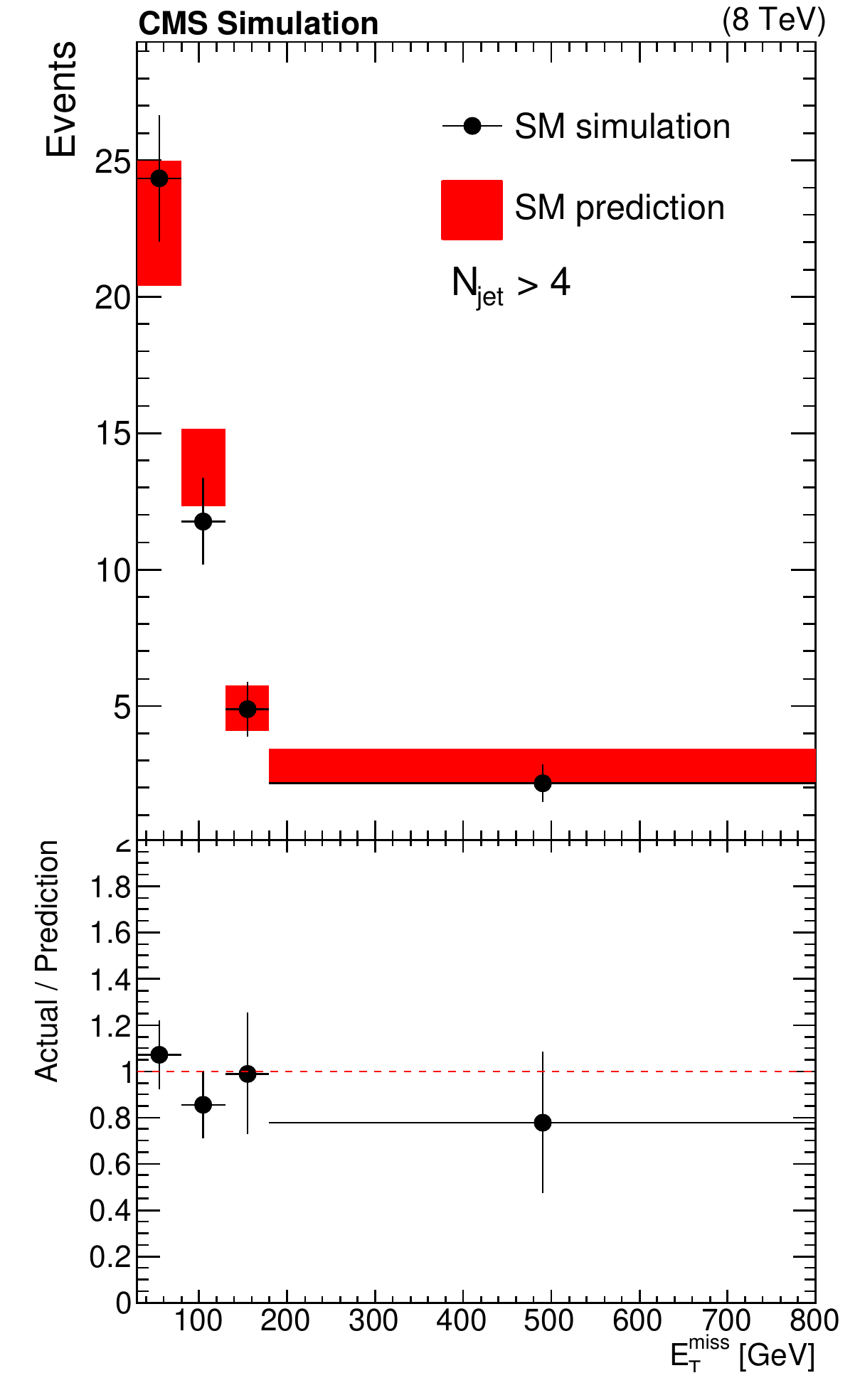}
      \caption{\ETmiss distribution predicted for SM backgrounds by extrapolation from the control region with the statistical and systematic uncertainties (red bands), compared to actual distribution (black points) for (left)~simulated events in the cross-check region, (middle)~data events in the cross-check region, and (right)~simulated events in the search region. The lower panels show the ratios of the actual to predicted distributions.}
      \label{fig9}
 \end{figure*}

The performance of the background estimation method is studied both in the SR, using simulation, and in a cross-check region defined by $2 \leq \njets \leq 4$, using data and simulation.  For both regions, the SM background consists primarily of \ttbar events, with a small contribution from \PW+jets production.
Figure~\ref{fig9} shows agreement between the predicted and actual \ETmiss distributions for the SR and cross-check regions.

The systematic uncertainty in the background prediction is based on the statistical uncertainties in the data, used to extract the $R_\text{ext}$ factors, and on the level of agreement between the predicted and actual results found using simulation in the SR (Fig.~\ref{fig9}\,right). No significant bias in the method is observed in simulation, and an additional systematic
uncertainty of 25--50\% is assigned to account for the statistical precision of the
  latter term.

\begin{figure}[htbp]
   \centering
       \includegraphics[width=\cmsFigWidth]{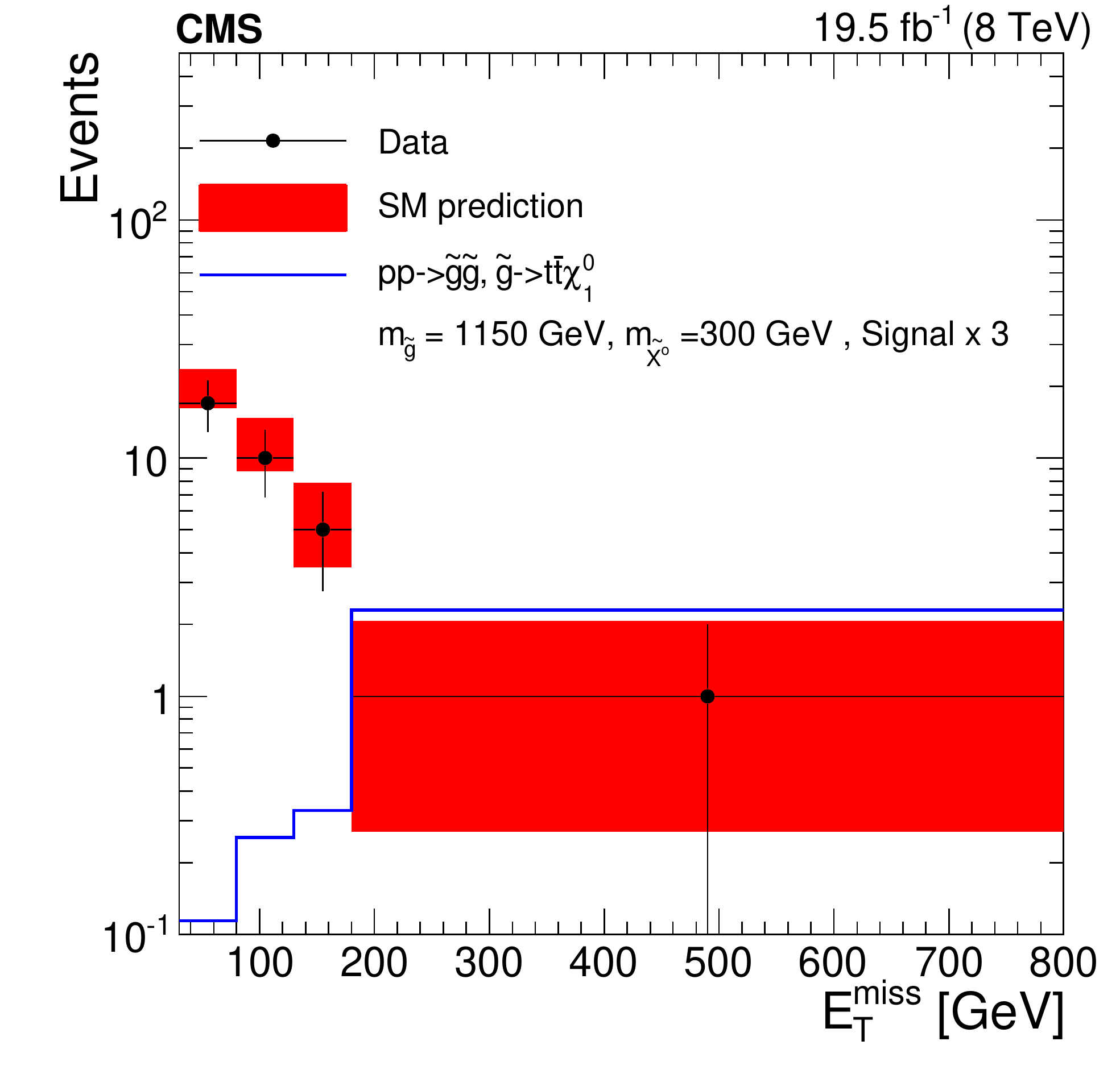}
       \caption{\ETmiss distribution in the signal region (black points) compared to the SM background prediction with its statistical and systematic uncertainties (red band). The expected signal for the T1tttt model with a gluino mass of 1150\GeV and an LSP mass of 300\GeV, multiplied by a factor of 3 for better visibility, is indicated by the blue curve.}
      \label{fig11}
 \end{figure}

The predicted and observed \ETmiss\ distributions for the signal region are shown in Fig.~\ref{fig11} and listed in Table~\ref{yields}. No excess of events is observed with respect to the SM prediction. For the interpretation of results (Section 5), all four \ETmiss bins are used. Besides their use in the combination, we present in \ifthenelse{\boolean{cms@external}}{}{Appendix} \ref{OSdileptonresults} the interpretation of the T1tttt and T5tttt scenarios  based on the results of the opposite-side dilepton analysis alone.

\begin{table*}[htb]
\centering
\topcaption{Predicted SM background and observed data yields as a function of $\ETmiss$ for the opposite-sign dilepton analysis.  The uncertainties in the total background predictions include both the statistical and the systematic components.}
\begin{tabular}{rcc}
 \multicolumn{1}{c}{$\ETmiss$ requirement} & Background prediction & Observed data yields \\ \hline
 $30<\ETmiss< 80$\GeV & 19.9 $\pm$ 3.7 & 17  \\
 $80<\ETmiss< 130$\GeV & 11.8 $\pm$ 3.0   & 10  \\
 $130<\ETmiss< 180$\GeV &  5.7 $\pm$ 2.2   & 5  \\
 $\ETmiss> 180$\GeV &   1.2 $\pm$ 1.1  & 1  \\
\end{tabular}
\label{yields}
\end{table*}

\section{Combination of analyses}\label{sec:Results}

The results from the five analyses are combined to provide more stringent conclusions.  The combined results are interpreted in the context of the SUSY scenarios illustrated in Fig.~\ref{fig:feyn}. The 95\% confidence level (CL) upper limits (UL) on the cross sections are calculated  using the LHC-style CL$_\mathrm{S}$ method~\cite{ref:STAT1,ref:STAT2,LHC-HCG}. Because of their large branching fractions, the fully hadronic and single-lepton analyses are most sensitive  in
 the largest part of the phase space. However, the analyses based on higher lepton
 multiplicities become important for the more compressed mass spectra and for models with fewer \bjets.

Systematic uncertainties in the signal selection efficiency are evaluated using the same techniques
 for all analyses.  They are  evaluated separately for the different signal models, search regions, and for each hypothesis for the SUSY particle masses.  The systematic uncertainties in the signal modeling are taken to be 100\% correlated among the mass hypotheses.  As an example, a summary of systematic uncertainties for  the T1tttt model is given in Table~\ref{tab:sigsyst}.  The total systematic uncertainty varies
 between 7 and 35\% depending on the decay modes considered, the search regions, and the mass points.  An important source of systematic uncertainty for the analyses that require multiple leptons arises from
  the lepton identification and isolation efficiencies, which are evaluated using $\cPZ\to\ell^+\ell^-$ events. The uncertainty in the energy scale of jets gives rise to a 1--15\% systematic uncertainty that increases with more stringent kinematic requirements.
 For compressed spectra, the modeling
 of initial-state radiation (ISR)~\cite{Chatrchyan:2013xna} is an important source of uncertainty. The PDF4LHC recommendations
 ~\cite{Alekhin:2011sk,ref:PDF4LHC} are used to evaluate the uncertainty associated with  the parton distribution functions (PDFs). For most of the analyses the background evaluation methods differ,
  and so the systematic uncertainties are treated as uncorrelated.  The overlap between most control regions is studied and found to be negligible.  The only exception occurs for the same-sign dilepton and multilepton analyses, which use the same methods to predict the background from non-prompt leptons and rare SM processes. For this case, the systematic uncertainties are taken to be fully correlated.

\begin{table*}[h]
\caption{Relative (\%) systematic uncertainties in the signal efficiency of the T1tttt model for the  fully hadronic (0~$\ell$), single-lepton  (1 $\ell$), opposite-sign dilepton (2 OS $\ell$),
  same-sign dilepton (2 SS $\ell$), and multilepton ($\ge$3 $\ell$) analyses. The given ranges reflect the variation across the different search  regions and for different values of the SUSY particle masses.}
\begin{center}
\begin{tabular}{ l   c c  c  c  c }
Source  &  0~$\ell$ &  1~$\ell$  & 2~OS~$\ell$ & 2~SS~$\ell$  & $\ge$3~$\ell$  \\ \hline
Integrated luminosity~\cite{LUMIPAS} &  \multicolumn{5}{c}{2.6}   \\
Pileup  &  \multicolumn{5}{c}{$<5$}   \\\cline{2-6}
Lepton identification and isolation efficiency & $<$1 & 3 & 4 & 10 & 12 \\
Trigger  efficiency & 2 & 4 & 6 & 6 & 5  \\
Parton distribution functions & 1--8 & 10--30 & 2 & 2 & 4 \\
Jet energy scale & 2--8 & 2--7 & 8 & 1--10 & 5--15\\
\cPqb-tagged jet identification & n/a & 1--15 & 14 & 2--10 & 5--20 \\
Initial-state radiation & 3--22 & 2--18 & 1--18 & 3--15 & 3--15 \\\hline
Total  & 7--28 & 14--35 & 17--25 & 18--25 & 15--30  \\
\end{tabular}
\label{tab:sigsyst}
\end{center}
\end{table*}

\subsection{Gluino-mediated top squark production with virtual top squarks}
\begin{figure}[htbp]
\centering
\includegraphics[width=0.49\textwidth]{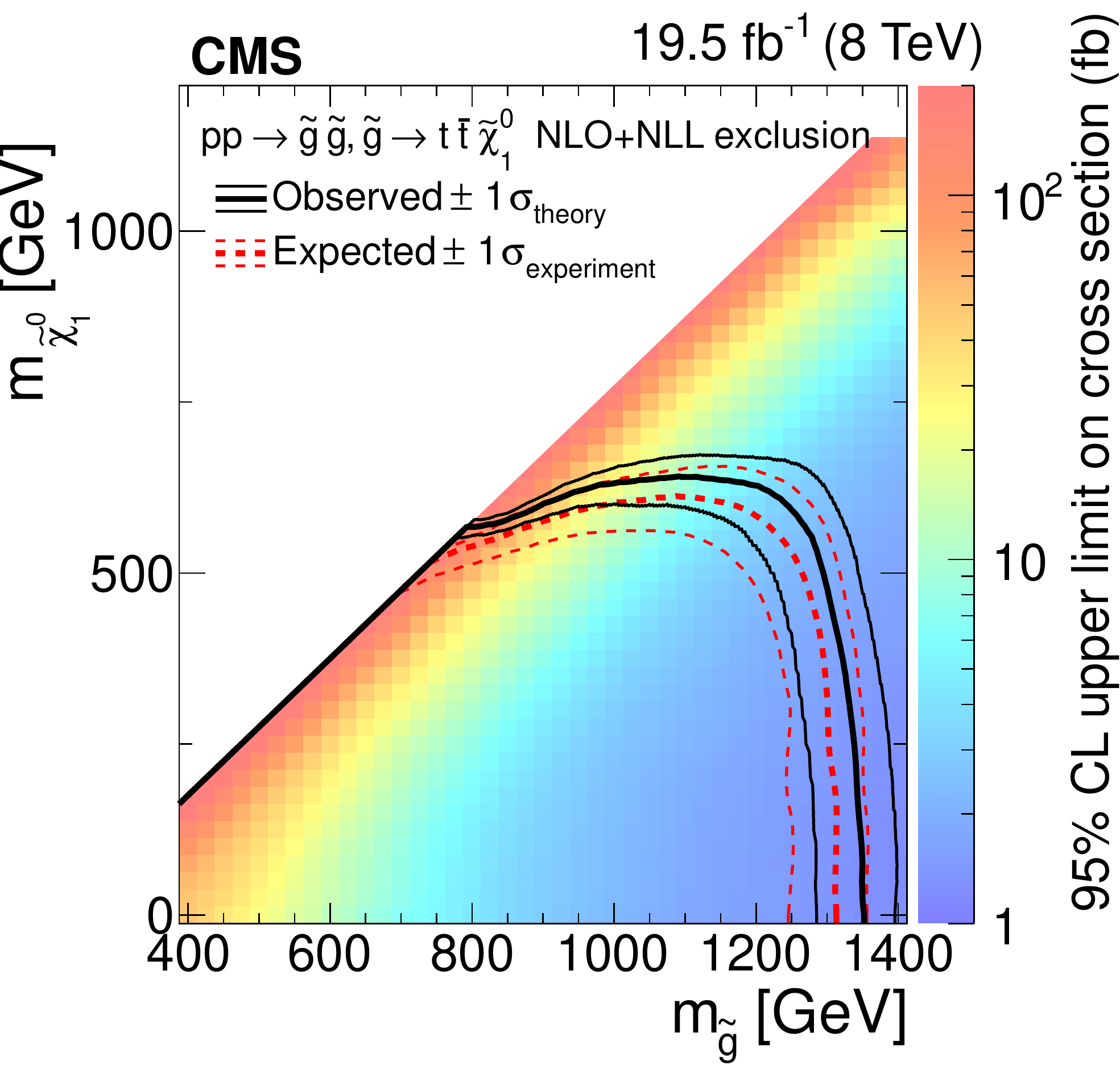}
\includegraphics[width=0.49\textwidth]{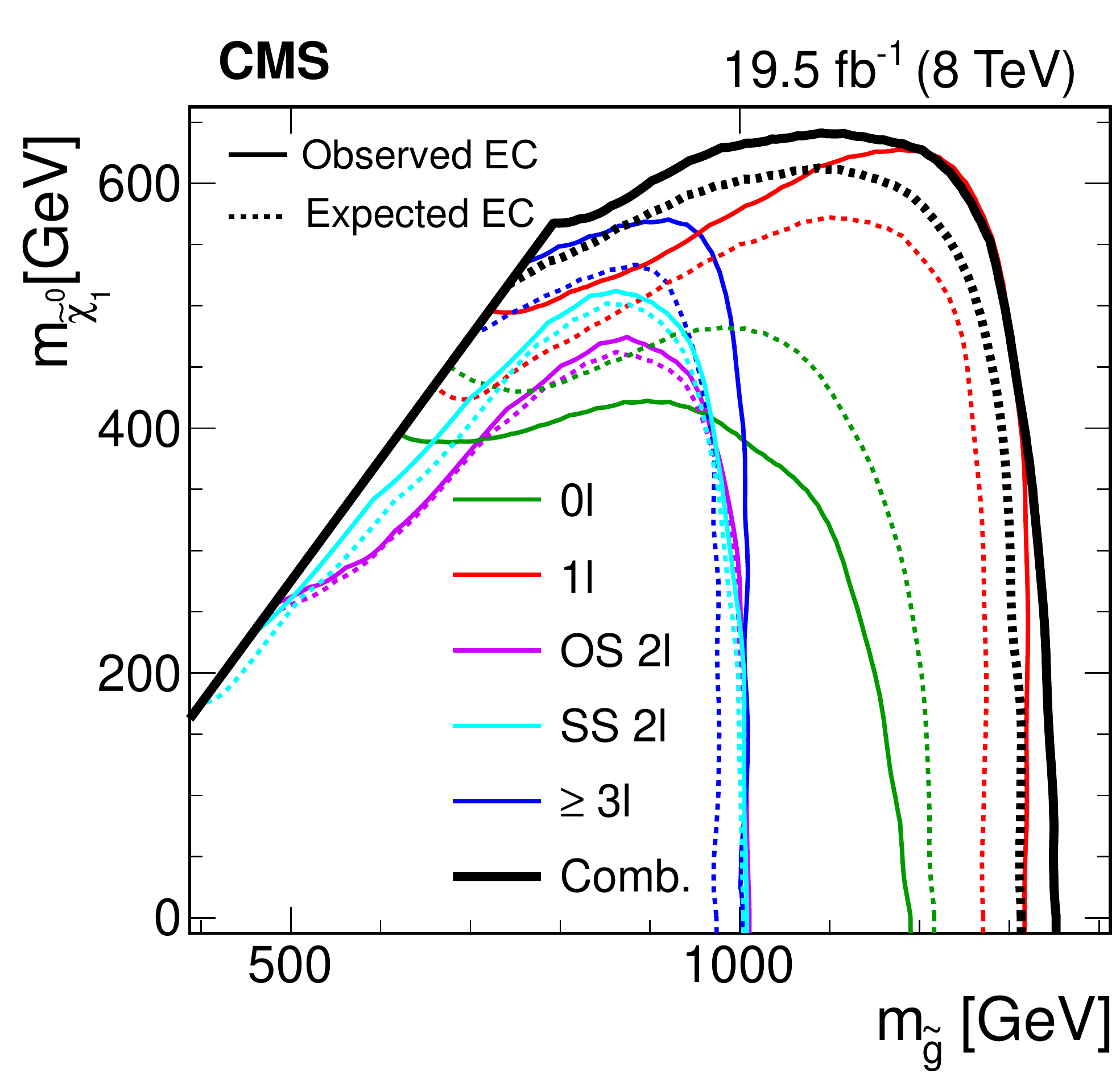}
\caption{(\cmsLeft) The 95\% CL cross section upper limits for gluino-mediated squark production with virtual top squarks, based on an NLO+NLL reference cross section for gluino pair production. The solid and dashed lines indicate, respectively, the observed and expected  exclusion contours for the combination of the five analyses.  The thin contours indicate the $\pm$1 standard deviation regions.  (\cmsRight) Exclusion contours (EC) for the individual searches, plus the combination.
\label{fig:T1ttttComb}
}
\end{figure}

The results are first interpreted in the context of $\sGlu\sGlu$ production with $\sGlu\to \ttbar
\chiz_1$ through a virtual $\sTop$, the process referred to as T1tttt.  The signature contains four top quarks and has significant jet activity (Fig.~\ref{fig:feyn}\,left).
The fully hadronic and single-lepton analyses are therefore expected to be especially sensitive, because of their larger signal efficiencies.
Figure~\ref{fig:T1ttttComb} (left) shows the 95\% CL upper limits on the product of the cross section and branching fraction in
the ($m_{\PSGczDo}$,$m_{\PSg}$)  plane.  The exclusion curves are evaluated by comparing the cross section upper limits with the next-to-leading-order (NLO) plus next-to-leading-logarithm (NLL) theoretical production cross sections~\cite{ref:SUSY_CROSS, ref:SUSY_CROSS2,ref:SUSY_CROSS3,ref:SUSY_CROSS4,ref:SUSY_CROSS5}.
The thick red dashed line indicates the 95\% CL expected limit, which is defined as the median of the upper limit distribution obtained using
pseudo-experiments and a likelihood model. The $\pm$1 standard deviation experimental systematic
  uncertainties $\sigma_{\text{experiment}}$ are shown by the thin red line around the expected limit.
The observed limit is given by the thick solid black line, where the uncertainty band (thin black lines) indicates the $\pm$1 standard deviation uncertainty $\sigma_{\text{theory}}$  in the theoretical
cross section. The theoretical uncertainty is mainly due to uncertainties in the renormalization and factorization scales, and in the knowledge of the PDFs. To quote the gluino mass exclusion, we conservatively consider the observed upper limit minus $\sigma_{\text{theory}}$.  It is seen that gluinos below 1280\GeV are excluded for $m_{\PSGczDo}\approx0$\GeV.  Assuming a gluino mass of 1000\GeV, an LSP with a mass below 600 \GeV is excluded.

The exclusion curves for each individual analysis are shown in Fig.~\ref{fig:T1ttttComb} (right). As expected, for low
LSP masses, the single-lepton and fully hadronic analyses provide the most stringent results.  For $m_{\PSGczDo}\approx0$\GeV, the combination is seen to extend the gluino mass exclusion by about 35\GeV compared to the single-lepton analysis, which provides the most stringent corresponding individual result.   Large values of  $m_{\PSGczDo}$ lead to more
compressed mass spectra, softer decay products, and therefore smaller \ETmiss.  As a result, the fully hadronic and single-lepton analyses become less sensitive,
since they require high-$\pt$ jets and large \ETmiss.
The dilepton and
multilepton analyses depend less on the $\pt$ spectrum of the final-state particles, and their sensitivity decreases less for smaller mass splittings. Thus the analyses requiring two or more leptons contribute most to the overall sensitivity when the difference $m_{\PSg}-m_{\PSGczDo}$ becomes small.   For $m_{\PSg}\approx1000$\GeV, the exclusion limit on $m_{\PSGczDo}$ is extended by about 60\GeV because of the addition of the multilepton channels.

\subsection{Gluino-mediated top squark production with on-shell top squarks}

\begin{figure}[htbp]
\centering
\includegraphics[width=0.49\textwidth]{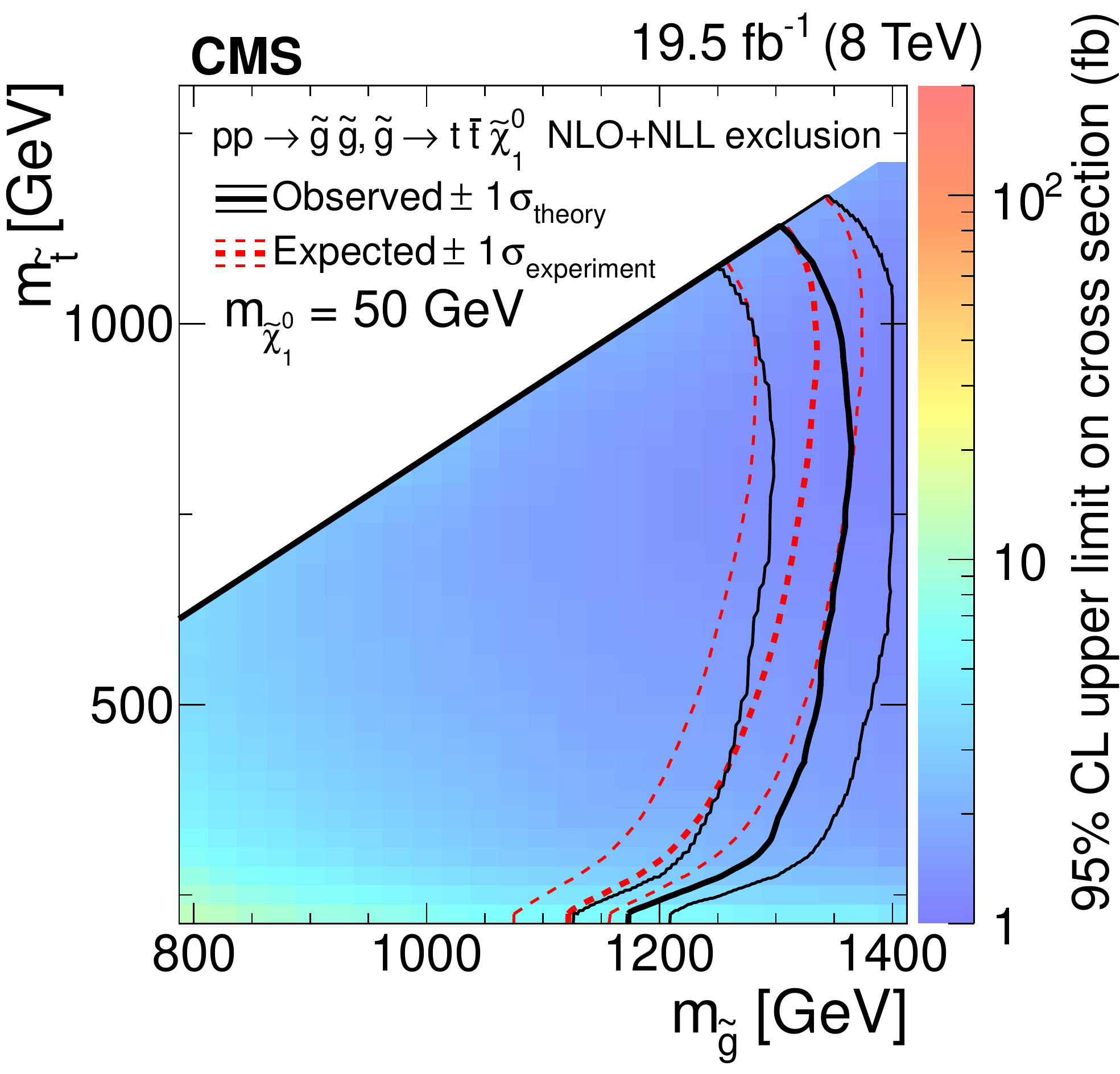}
\includegraphics[width=0.49\textwidth]{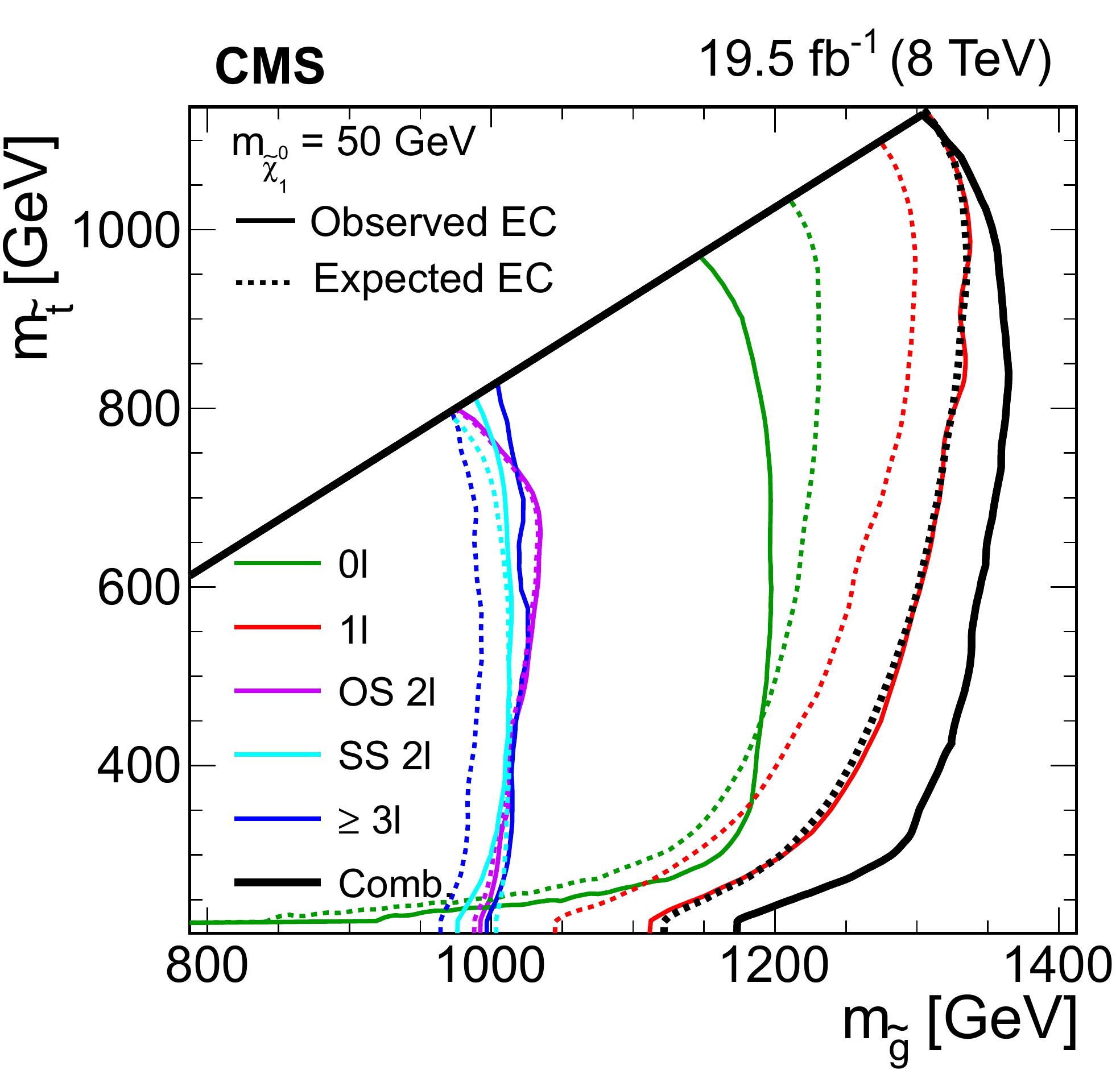} \caption{(\cmsLeft)~The 95\% CL cross section upper limits for gluino-mediated squark production with on-shell top squarks, assuming an LSP mass of $m_{\PSGczDo}=$50\GeV, based on an NLO+NLL reference cross section for gluino pair production. The solid and dashed lines indicate, respectively, the observed and expected exclusion contours for the combination of the five analyses. The thin contours indicate the $\pm$1 standard deviation regions.
(\cmsRight)~Exclusion contours (EC) for the individual searches, plus the combination.
\label{fig:T5ttttComb}
}
\end{figure}

If the top squarks are light enough, the gluino can decay through an intermediate on-shell top squark.  In this
model, referred to as T5tttt, the values of $m_{\PSGczDo}$, $m_{\PSQt}$, and $m_{\PSg}$ function as independent parameters.
Results are presented for a fixed mass $m_{\PSGczDo}=$50\GeV and scanned over the masses of the on-shell top squark and gluino.  The 95\% CL upper
limits on the product of the cross section and the branching fraction in the $m_{\PSQt}$ versus $m_{\PSg}$ plane are shown
in Fig.~\ref{fig:T5ttttComb}, \cmsLeft. In the context of the T5tttt model, gluinos with masses below around 1300\GeV are excluded for top squark masses around 700\GeV.  Figure~\ref{fig:T5ttttComb}, \cmsRight, shows the results for the individual studies.  The contribution from the fully hadronic analysis remains important even for relatively small top squark masses $m_{\PSQt}\approx 200\gev$ because of the high \HT search regions: signal events in this case contain smaller \ETmiss but larger \HT.  However, for $m_{\PSQt}< 150\GeV$, the fully hadronic analysis loses sensitivity.  The single-lepton analysis provides the most stringent individual results, but loses sensitivity as $m_{\PSQt}$ decreases.  The sensitivity of the dilepton and multilepton searches depends less strongly on $m_{\PSQt}$, but
their sensitivity even in the compressed region is rather small, although they contribute to the combination at very small $m_{\PSQt}$.  The combination improves the exclusion reach in the gluino mass by about 50\GeV for small $m_{\PSQt}$.

\subsection{Bottom squark pair production}

We also consider bottom squark pair production with the bottom squarks decaying as $\sBot_1 \to \cPqt \chim_1$ and $\chim_1 \to \PWm \chiz_1$,
known as T6ttWW (Fig.~\ref{fig:feyn}\,right). The single-lepton and opposite-sign dilepton analyses have very little sensitivity to such
a model because of the stringent $\nbjets$ and $\njets$  requirements, and are not included in the combination.  The fully hadronic analysis, which does not impose a requirement on $\nbjets$, contributes some sensitivity. The main sensitivity comes from the same-sign and multilepton searches with either $\nbjets= 1$ or 2.

\begin{figure}[htbp]
\centering
\includegraphics[width=0.49\textwidth]{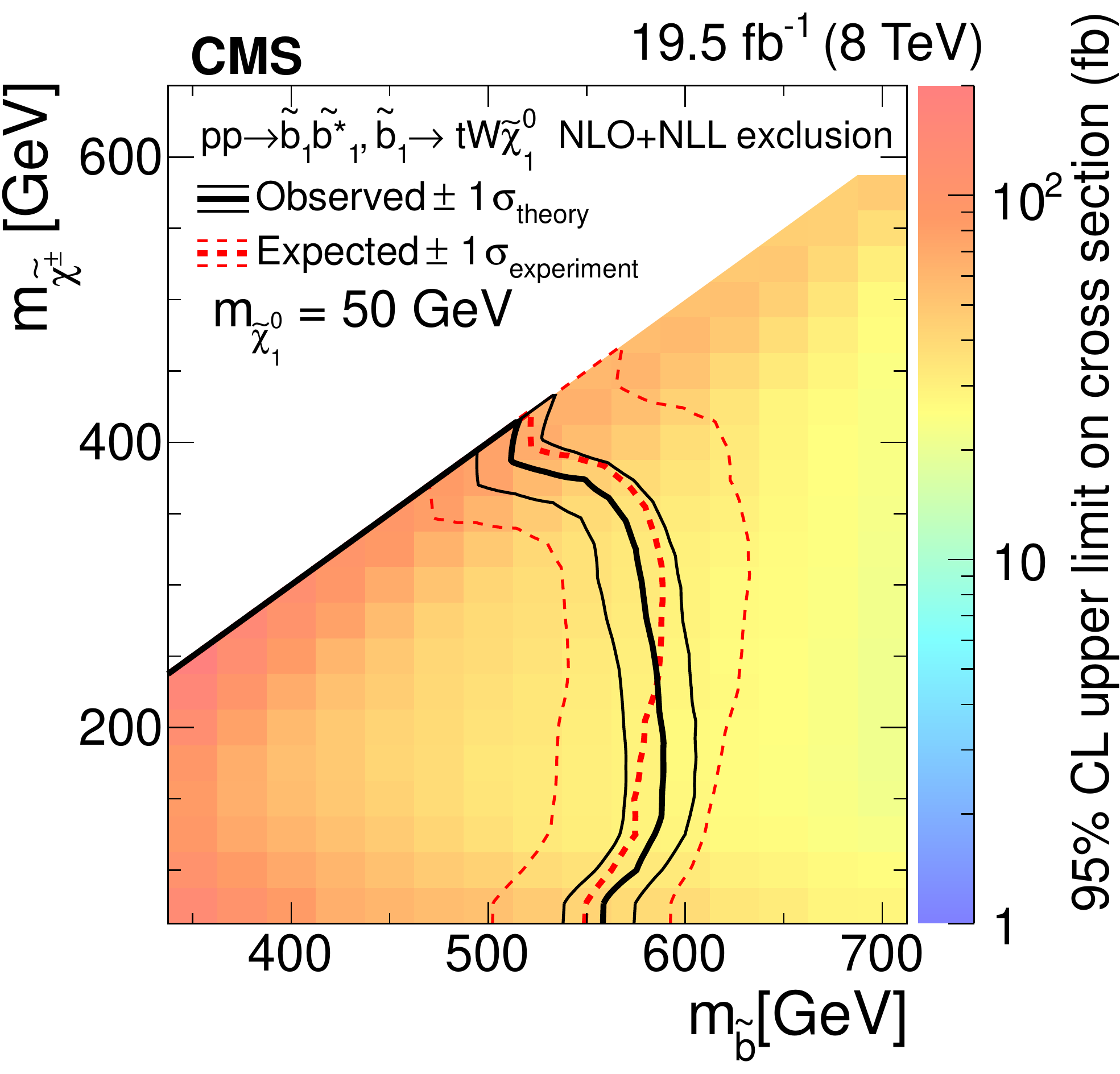}
\includegraphics[width=0.49\textwidth]{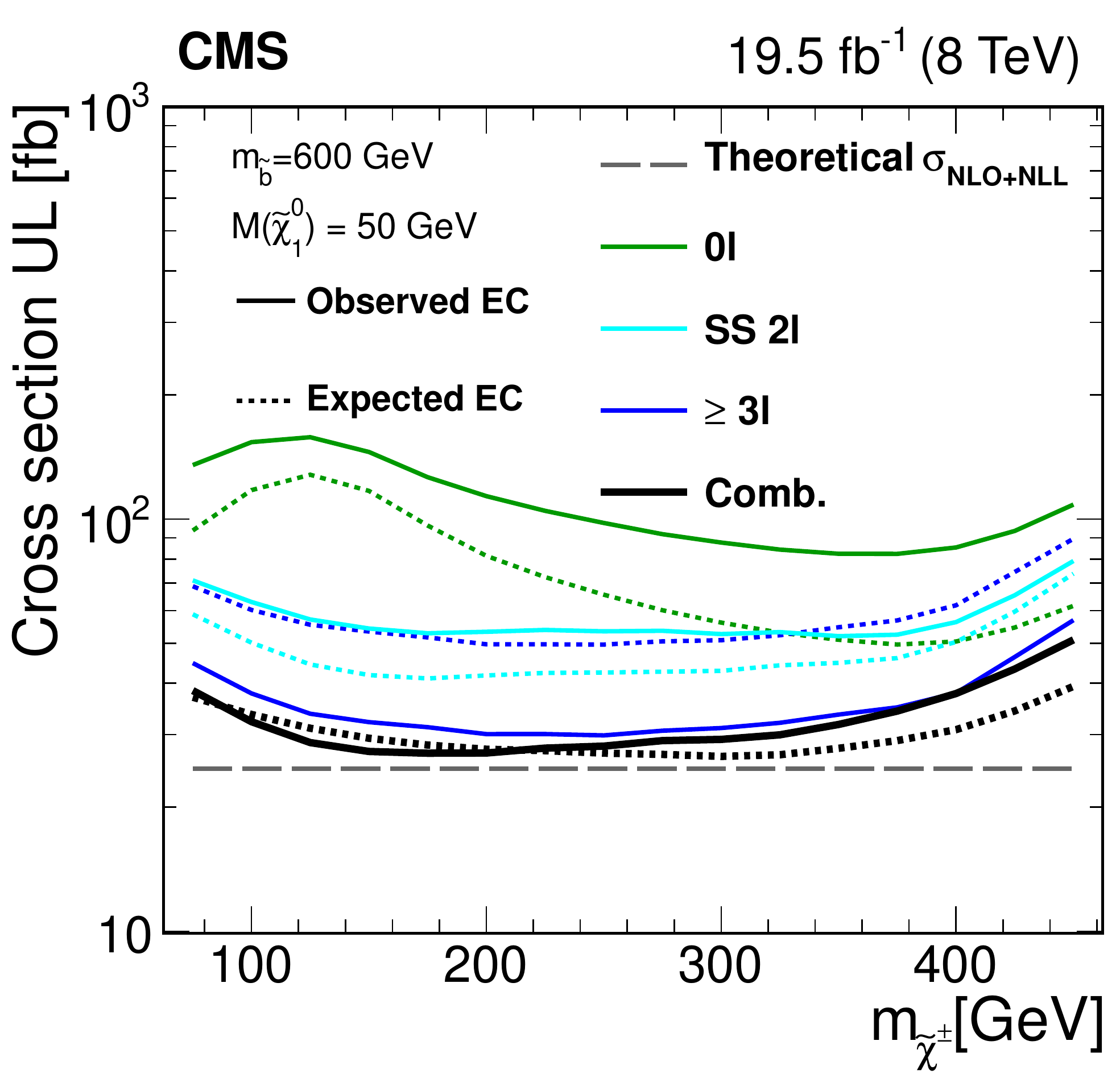}
\caption{(\cmsLeft) The 95\% CL cross section upper limits for bottom-squark pair production, assuming an LSP  mass of $m_{\PSGczDo}=$50\GeV, based on an NLO+NLL reference cross section. The solid and dashed lines indicate, respectively, the observed and expected  exclusion contours for the combination of the fully hadronic,  same-sign dilepton, and multilepton analyses.  The thin contours indicate the $\pm$1 standard deviation regions.  (\cmsRight) Exclusion contours (EC) for the individual searches, plus the combination, assuming a bottom squark mass of 600\GeV. \label{fig:T6ttWWComb}
}
\end{figure}

For the T6ttWW model, the LSP mass is set to 50\GeV. The resulting 95\% CL upper limits on the product of the cross section and branching fraction in the $m_{\PSGcpm}$  versus $m_{\PSQb}$ plane are shown in Fig.~\ref{fig:T6ttWWComb},\,\cmsLeft.  In the context of this model, bottom squark masses up to 570\GeV are excluded for LSP masses around 150-300\GeV.  Figure~\ref{fig:T6ttWWComb},\,\cmsRight, shows the exclusion limits for the individual analyses assuming a fixed bottom squark mass of 600\GeV.  The same-sign dilepton analysis
provides the best sensitivity for chargino masses below 400\GeV, and the combination with the multilepton analysis leads to a 15\% improvement in the cross section upper limit and even up to 35\% improvement in the expected cross section upper limit, which represents an improvement in the expected sbottom mass exclusion limits of around 50\GeV. For larger chargino masses, the fully hadronic analysis is more sensitive because jets from \PW\ boson decays become more energetic.

\section{Summary}
Five searches for supersymmetry with non-overlapping event samples are combined to obtain
more stringent exclusion limits on models in which \bjets and four \PW\ bosons are produced.  The results are based on data corresponding to an integrated luminosity of 19.5\fbinv of pp collisions, collected with the CMS detector at $\sqrt{s}=8\TeV$ in 2012.  Because of their large branching fractions, the single-lepton and fully hadronic analyses have the largest sensitivity for most of the range of the supersymmetric mass spectra, whereas the analyses with higher lepton multiplicities have higher sensitivity for models with more compressed mass spectra. The complementarity of the searches is exploited to provide comprehensive coverage across a wide region of parameter space.  The combined searches yield 95\% confidence level exclusions of up to 1280 and 570\GeV for the gluino and bottom-squark masses in the context of gluino and bottom-squark pair production, respectively. The increase in sensitivity that arises from the combination of the five analyses corresponds to an increase of about 50\GeV in the SUSY mass reach compared to the individual analyses.

\begin{acknowledgments}
We congratulate our colleagues in the CERN accelerator departments for the excellent performance of the LHC and thank the technical and administrative staffs at CERN and at other CMS institutes for their contributions to the success of the CMS effort. In addition, we gratefully acknowledge the computing centers and personnel of the Worldwide LHC Computing Grid for delivering so effectively the computing infrastructure essential to our analyses. Finally, we acknowledge the enduring support for the construction and operation of the LHC and the CMS detector provided by the following funding agencies: BMWFW and FWF (Austria); FNRS and FWO (Belgium); CNPq, CAPES, FAPERJ, and FAPESP (Brazil); MES (Bulgaria); CERN; CAS, MoST, and NSFC (China); COLCIENCIAS (Colombia); MSES and CSF (Croatia); RPF (Cyprus); MoER, ERC IUT and ERDF (Estonia); Academy of Finland, MEC, and HIP (Finland); CEA and CNRS/IN2P3 (France); BMBF, DFG, and HGF (Germany); GSRT (Greece); OTKA and NIH (Hungary); DAE and DST (India); IPM (Iran); SFI (Ireland); INFN (Italy); MSIP and NRF (Republic of Korea); LAS (Lithuania); MOE and UM (Malaysia); CINVESTAV, CONACYT, SEP, and UASLP-FAI (Mexico); MBIE (New Zealand); PAEC (Pakistan); MSHE and NSC (Poland); FCT (Portugal); JINR (Dubna); MON, RosAtom, RAS and RFBR (Russia); MESTD (Serbia); SEIDI and CPAN (Spain); Swiss Funding Agencies (Switzerland); MST (Taipei); ThEPCenter, IPST, STAR and NSTDA (Thailand); TUBITAK and TAEK (Turkey); NASU and SFFR (Ukraine); STFC (United Kingdom); DOE and NSF (USA).

Individuals have received support from the Marie-Curie programme and the European Research Council and EPLANET (European Union); the Leventis Foundation; the A. P. Sloan Foundation; the Alexander von Humboldt Foundation; the Belgian Federal Science Policy Office; the Fonds pour la Formation \`a la Recherche dans l'Industrie et dans l'Agriculture (FRIA-Belgium); the Agentschap voor Innovatie door Wetenschap en Technologie (IWT-Belgium); the Ministry of Education, Youth and Sports (MEYS) of the Czech Republic; the Council of Science and Industrial Research, India; the HOMING PLUS programme of Foundation for Polish Science, cofinanced from European Union, Regional Development Fund; the Compagnia di San Paolo (Torino); the Consorzio per la Fisica (Trieste); MIUR project 20108T4XTM (Italy); the Thalis and Aristeia programmes cofinanced by EU-ESF and the Greek NSRF; and the National Priorities Research Program by Qatar National Research Fund.
\end{acknowledgments}
\bibliography{auto_generated}   

\appendix
\section{Additional plots for the opposite-sign dilepton search \label{OSdileptonresults}}
\begin{figure}[tbhp]
\centering
\includegraphics[width=0.48\textwidth]{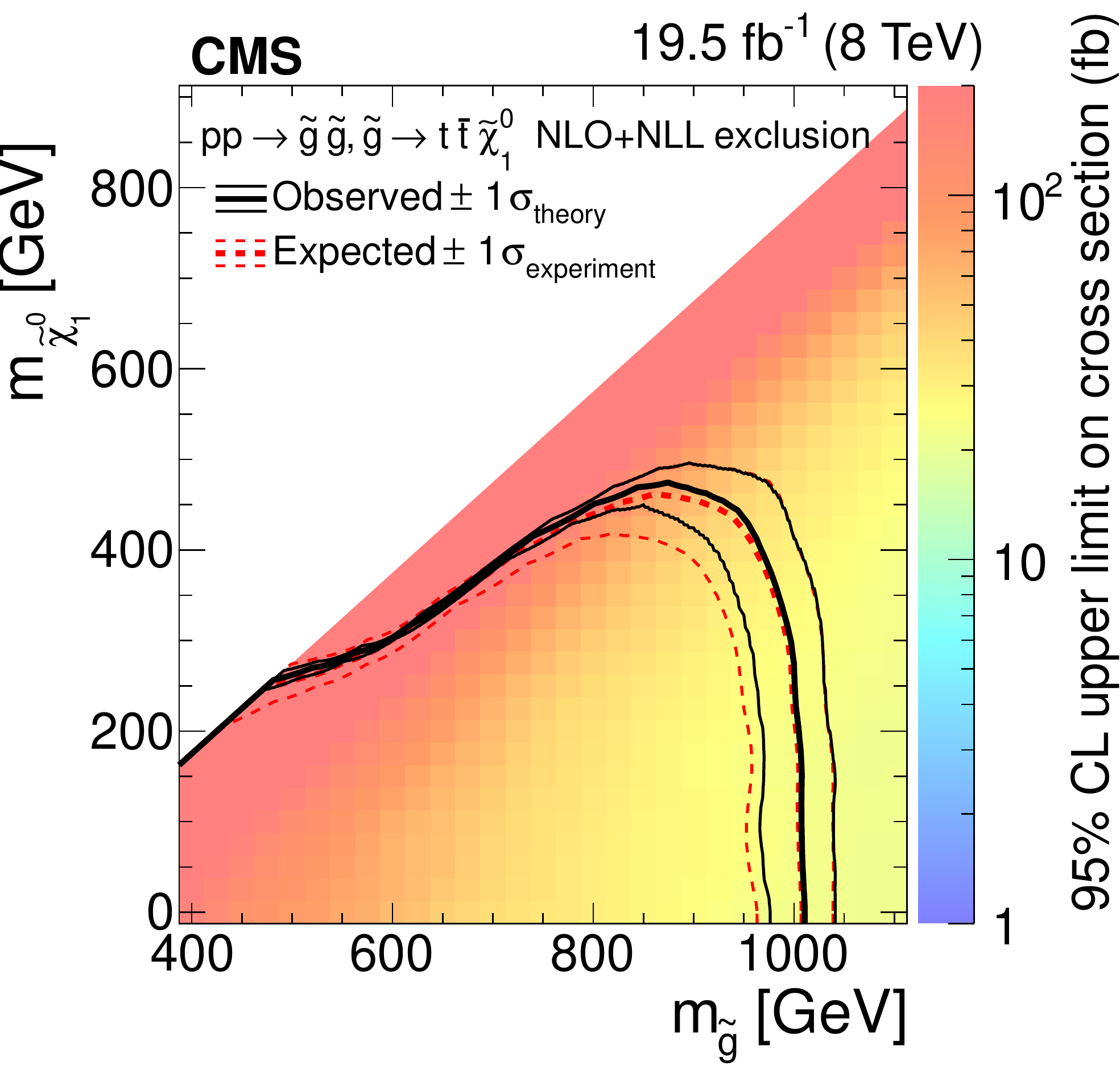}
\includegraphics[width=0.48\textwidth]{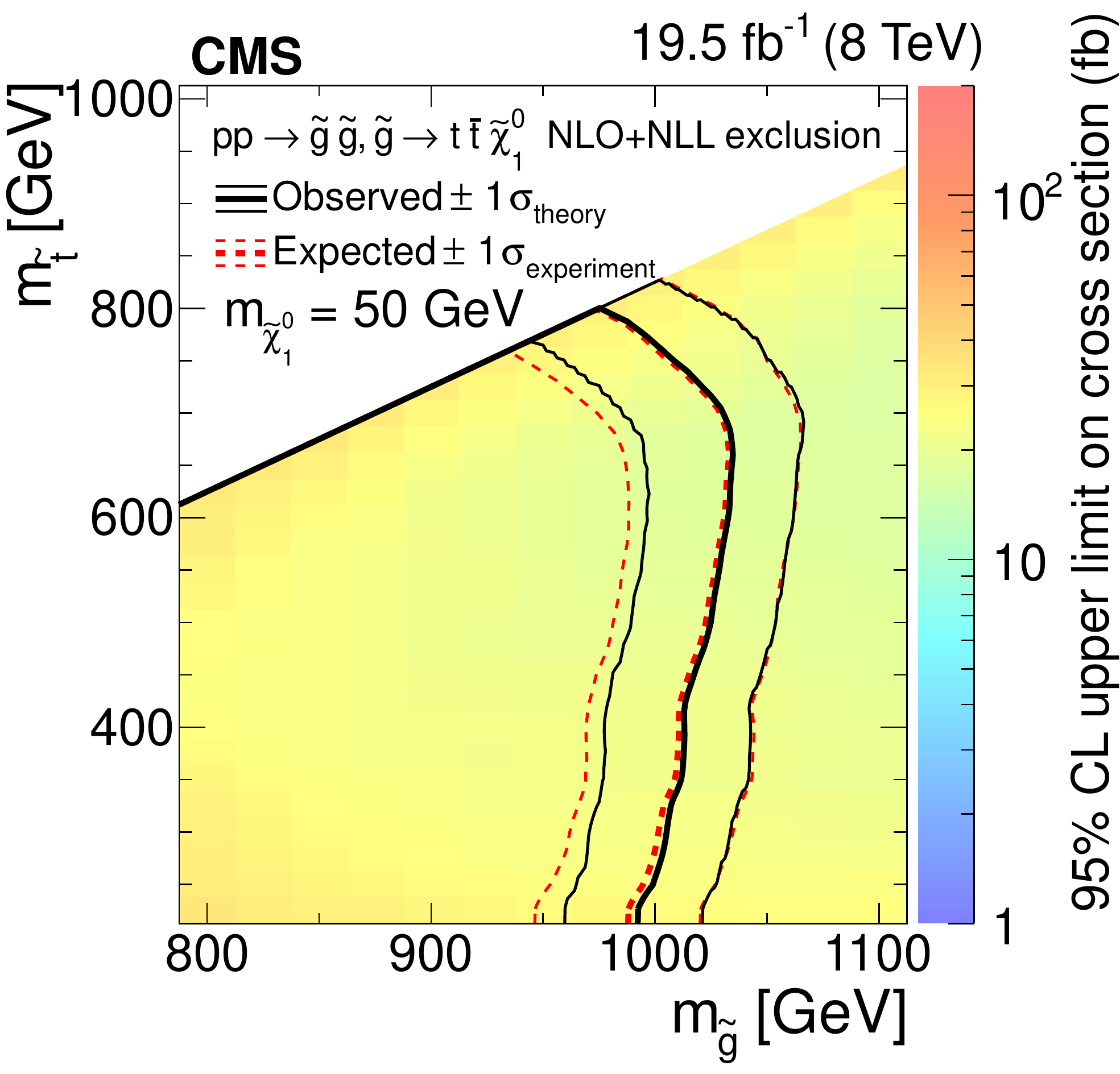}
\caption{(\cmsLeft) The 95\% CL cross section upper limits for gluino-mediated squark production with virtual top squarks, based on an NLO+NLL reference cross section for gluino pair production, derived from the opposite-sign dilepton analysis. The solid and dashed lines indicate, respectively, the observed and expected exclusion contours. The thin contours indicate the $\pm$1 standard deviation regions. (\cmsRight) The corresponding results for gluino-mediated squark production with on-shell top squarks.
\label{fig:interpretationOS2l}
}
\end{figure}

This appendix presents additional results for the opposite-sign dilepton search.  The results of this analysis alone for the T1tttt (Fig.~\ref{fig:feyn}\,left) and T5tttt (Fig.~\ref{fig:feyn}\,middle) models are shown, respectively, in Figs.~\ref{fig:interpretationOS2l}, \cmsLeft and \cmsRight.  In the context of the T1tttt model, gluinos with masses below around 980\GeV are excluded for LSP masses below 400\GeV.  In the T5tttt model, gluinos with masses below 1000\GeV are probed for top squark masses around 650\GeV.

\cleardoublepage \section{The CMS Collaboration \label{app:collab}}\begin{sloppypar}\hyphenpenalty=5000\widowpenalty=500\clubpenalty=5000\input{SUS-14-010-authorlist.tex}\end{sloppypar}
\end{document}

%% file: SUS-14-010-authorlist.tex
\textbf{Yerevan Physics Institute,  Yerevan,  Armenia}\\*[0pt]
V.~Khachatryan, A.M.~Sirunyan, A.~Tumasyan
\vskip\cmsinstskip
\textbf{Institut f\"{u}r Hochenergiephysik der OeAW,  Wien,  Austria}\\*[0pt]
W.~Adam, T.~Bergauer, M.~Dragicevic, J.~Er\"{o}, M.~Friedl, R.~Fr\"{u}hwirth\cmsAuthorMark{1}, V.M.~Ghete, C.~Hartl, N.~H\"{o}rmann, J.~Hrubec, M.~Jeitler\cmsAuthorMark{1}, W.~Kiesenhofer, V.~Kn\"{u}nz, M.~Krammer\cmsAuthorMark{1}, I.~Kr\"{a}tschmer, D.~Liko, I.~Mikulec, D.~Rabady\cmsAuthorMark{2}, B.~Rahbaran, H.~Rohringer, R.~Sch\"{o}fbeck, J.~Strauss, W.~Treberer-Treberspurg, W.~Waltenberger, C.-E.~Wulz\cmsAuthorMark{1}
% ...
\vskip\cmsinstskip
% ...
1:~~Also at Vienna University of Technology, Vienna, Austria\\
2:~~Also at CERN, European Organization for Nuclear Research, Geneva, Switzerland\\
% ...


%% file: SUS-14-010_temp.bbl
\providecommand{\href}[2]{#2}\begingroup\raggedright\begin{thebibliography}{10}%
\makeatletter
\providecommand{\hrefCMSnoop }[0]{\@secondoftwo}%
\makeatother
\providecommand{\doi}{\texttt{doi:}\begingroup \urlstyle{tt}\Url}

\bibitem{Aad:2012tfa}
\hrefCMSnoop {}{{ATLAS} Collaboration, ``{Observation of a new particle in the
  search for the Standard Model Higgs boson with the ATLAS detector at the
  LHC}'',} \textit{ Phys. Lett. B} \textbf{ 716} (2012) 1,
  \href{http://dx.doi.org/10.1016/j.physletb.2012.08.020}{\doi{10.1016/j.physletb.2012.08.020}},
\href{http://www.arXiv.org/abs/1207.7214}{\texttt{arXiv:1207.7214}}.

\bibitem{Chatrchyan:2012ufa}
\hrefCMSnoop {}{{CMS} Collaboration, ``{Observation of a new boson at a mass of
  125 GeV with the CMS experiment at the LHC}'',} \textit{ Phys. Lett. B}
  \textbf{ 716} (2012) 30,
  \href{http://dx.doi.org/10.1016/j.physletb.2012.08.021}{\doi{10.1016/j.physletb.2012.08.021}},
\href{http://www.arXiv.org/abs/1207.7235}{\texttt{arXiv:1207.7235}}.

\bibitem{Barbieri:1987fn}
\hrefCMSnoop {}{R.~Barbieri and G.~F. Giudice, ``Upper bounds on supersymmetric
  particle masses'',} \textit{ Nucl. Phys. B} \textbf{ 306} (1988) 63,
\href{http://dx.doi.org/10.1016/0550-3213(88)90171-X}{\doi{10.1016/0550-3213(88)90171-X}}.

\bibitem{Romanino:1999ut}
\hrefCMSnoop {}{A.~Romanino and A.~Strumia, ``{Are heavy scalars natural in
  minimal supergravity?}'',} \textit{ Phys. Lett. B} \textbf{ 487} (2000) 165,
  \href{http://dx.doi.org/10.1016/S0370-2693(00)00806-6}{\doi{10.1016/S0370-2693(00)00806-6}},
\href{http://www.arXiv.org/abs/hep-ph/9912301}{\texttt{arXiv:hep-ph/9912301}}.

\bibitem{Feng:1999mn}
\hrefCMSnoop {}{J.~L. Feng, K.~T. Matchev, and T.~Moroi, ``Multi-{TeV} Scalars
  are Natural in Minimal Supergravity'',} \textit{ Phys. Rev. Lett.} \textbf{
  84} (2000) 2322,
  \href{http://dx.doi.org/10.1103/PhysRevLett.84.2322}{\doi{10.1103/PhysRevLett.84.2322}},
\href{http://www.arXiv.org/abs/hep-ph/9908309}{\texttt{arXiv:hep-ph/9908309}}.

\bibitem{Kitano:2005wc}
\hrefCMSnoop {}{R.~Kitano and Y.~Nomura, ``A solution to the supersymmetric
  fine-tuning problem within the {MSSM}'',} \textit{ Phys. Lett. B} \textbf{
  631} (2005) 58,
  \href{http://dx.doi.org/10.1016/j.physletb.2005.10.003}{\doi{10.1016/j.physletb.2005.10.003}},
\href{http://www.arXiv.org/abs/hep-ph/0509039}{\texttt{arXiv:hep-ph/0509039}}.

\bibitem{Giudice:2006sn}
\hrefCMSnoop {}{G.~F. Giudice and R.~Rattazzi, ``Living dangerously with
  low-energy supersymmetry'',} \textit{ Nucl. Phys. B} \textbf{ 757} (2006) 19,
  \href{http://dx.doi.org/10.1016/j.nuclphysb.2006.07.031}{\doi{10.1016/j.nuclphysb.2006.07.031}},
\href{http://www.arXiv.org/abs/hep-ph/0606105}{\texttt{arXiv:hep-ph/0606105}}.

\bibitem{Barbieri:2009ev}
\hrefCMSnoop {}{R.~Barbieri and D.~Pappadopulo, ``{S-particles at their
  naturalness limits}'',} \textit{ JHEP} \textbf{ 10} (2009) 061,
  \href{http://dx.doi.org/10.1088/1126-6708/2009/10/061}{\doi{10.1088/1126-6708/2009/10/061}},
\href{http://www.arXiv.org/abs/0906.4546}{\texttt{arXiv:0906.4546}}.

\bibitem{Horton:2009ed}
\hrefCMSnoop {}{D.~Horton and G.~G. Ross, ``Naturalness and focus points with
  non-universal gaugino masses'',} \textit{ Nucl. Phys. B} \textbf{ 830} (2010)
  221,
  \href{http://dx.doi.org/10.1016/j.nuclphysb.2009.12.031}{\doi{10.1016/j.nuclphysb.2009.12.031}},
\href{http://www.arXiv.org/abs/0908.0857}{\texttt{arXiv:0908.0857}}.

\bibitem{Farrar:1978xj}
\hrefCMSnoop {}{G.~R. Farrar and P.~Fayet, ``{Phenomenology of the production,
  decay, and detection of new hadronic states associated with
  supersymmetry}'',} \textit{ Phys. Lett. B} \textbf{ 76} (1978) 575,
\href{http://dx.doi.org/10.1016/0370-2693(78)90858-4}{\doi{10.1016/0370-2693(78)90858-4}}.

\bibitem{Aad:2013wta}
\hrefCMSnoop {}{{ATLAS} Collaboration, ``{Search for new phenomena in final
  states with large jet multiplicities and missing transverse momentum at
  $\sqrt{s}=8$~TeV proton-proton collisions using the ATLAS experiment}'',}
  \textit{ JHEP} \textbf{ 10} (2013) 130,
  \href{http://dx.doi.org/10.1007/JHEP10(2013)130}{\doi{10.1007/JHEP10(2013)130}},
\href{http://www.arXiv.org/abs/1308.1841}{\texttt{arXiv:1308.1841}}.

\bibitem{Aad:2013ija}
\hrefCMSnoop {}{{ATLAS} Collaboration, ``{Search for direct third-generation
  squark pair production in final states with missing transverse momentum and
  two \b jets in $\sqrt{s} =$ 8~TeV pp collisions with the ATLAS detector}'',}
  \textit{ JHEP} \textbf{ 10} (2013) 189,
  \href{http://dx.doi.org/10.1007/JHEP10(2013)189}{\doi{10.1007/JHEP10(2013)189}},
\href{http://www.arXiv.org/abs/1308.2631}{\texttt{arXiv:1308.2631}}.

\bibitem{Aad:2014nua}
\hrefCMSnoop {}{{ATLAS} Collaboration, ``{Search for direct production of
  charginos and neutralinos in events with three leptons and missing transverse
  momentum in $\sqrt{s}=8$~TeV pp collisions with the ATLAS detector}'',}
  \textit{ JHEP} \textbf{ 04} (2014) 169,
  \href{http://dx.doi.org/10.1007/JHEP04(2014)169}{\doi{10.1007/JHEP04(2014)169}},
\href{http://www.arXiv.org/abs/1402.7029}{\texttt{arXiv:1402.7029}}.

\bibitem{Aad:2014qaa}
\hrefCMSnoop {}{{ATLAS} Collaboration, ``{Search for direct top-squark pair
  production in final states with two leptons in pp collisions at
  $\sqrt{s}=8$~TeV with the ATLAS detector}'',} \textit{ JHEP} \textbf{ 06}
  (2014) 124,
  \href{http://dx.doi.org/10.1007/JHEP06(2014)124}{\doi{10.1007/JHEP06(2014)124}},
\href{http://www.arXiv.org/abs/1403.4853}{\texttt{arXiv:1403.4853}}.

\bibitem{Aad:2014vma}
\hrefCMSnoop {}{{ATLAS} Collaboration, ``{Search for direct production of
  charginos, neutralinos and sleptons in final states with two leptons and
  missing transverse momentum in pp collisions at $\sqrt{s}= 8$~TeV with the
  ATLAS detector}'',} \textit{ JHEP} \textbf{ 05} (2014) 071,
  \href{http://dx.doi.org/10.1007/JHEP05(2014)071}{\doi{10.1007/JHEP05(2014)071}},
\href{http://www.arXiv.org/abs/1403.5294}{\texttt{arXiv:1403.5294}}.

\bibitem{Aad:2014iza}
\hrefCMSnoop {}{{ATLAS} Collaboration, ``{Search for supersymmetry in events
  with four or more leptons in $\sqrt{s}=8$~TeV pp collisions with the ATLAS
  detector}'',} (2014).
  \href{http://www.arXiv.org/abs/1405.5086}{\texttt{arXiv:1405.5086}}.
Submitted to {\it{Phys.\ Rev.\ D}}.

\bibitem{Chatrchyan:2013wxa}
\hrefCMSnoop {}{{CMS} Collaboration, ``{Search for gluino mediated bottom- and
  top-squark production in multijet final states in pp collisions at 8~TeV}'',}
  \textit{ Phys. Lett. B} \textbf{ 725} (2013) 243,
  \href{http://dx.doi.org/10.1016/j.physletb.2013.06.058}{\doi{10.1016/j.physletb.2013.06.058}},
\href{http://www.arXiv.org/abs/1305.2390}{\texttt{arXiv:1305.2390}}.

\bibitem{Chatrchyan:2013xna}
\hrefCMSnoop {}{{CMS} Collaboration, ``{Search for top-squark pair production
  in the single-lepton final state in pp collisions at $\sqrt{s}=8$ TeV}'',}
  \textit{ Eur. Phys. J. C} \textbf{ 73} (2013) 2677,
  \href{http://dx.doi.org/10.1140/epjc/s10052-013-2677-2}{\doi{10.1140/epjc/s10052-013-2677-2}},
\href{http://www.arXiv.org/abs/1308.1586}{\texttt{arXiv:1308.1586}}.

\bibitem{Chatrchyan:2014lfa}
\hrefCMSnoop {}{{CMS} Collaboration, ``{Search for new physics in the multijet
  and missing transverse momentum final state in proton-proton collisions at
  $\sqrt{s}=8$~TeV}'',} \textit{ JHEP} \textbf{ 06} (2014) 055,
  \href{http://dx.doi.org/10.1007/JHEP06(2014)055}{\doi{10.1007/JHEP06(2014)055}},
\href{http://www.arXiv.org/abs/1402.4770}{\texttt{arXiv:1402.4770}}.

\bibitem{Chatrchyan:2013iqa}
\hrefCMSnoop {}{{CMS} Collaboration, ``{Search for supersymmetry in pp
  collisions at $\sqrt{s}$ = 8 TeV in events with a single lepton, large jet
  multiplicity, and multiple b jets}'',} \textit{ Phys. Lett. B} \textbf{ 733}
  (2013) 328,
  \href{http://dx.doi.org/10.1016/j.physletb.2014.04.023}{\doi{10.1016/j.physletb.2014.04.023}},
\href{http://www.arXiv.org/abs/1311.4937}{\texttt{arXiv:1311.4937}}.

\bibitem{Chatrchyan:2013fea}
\hrefCMSnoop {}{{CMS} Collaboration, ``{Search for new physics in events with
  same-sign dileptons and jets in pp collisions at $\sqrt{s}$ = 8 TeV}'',}
  \textit{ JHEP} \textbf{ 01} (2014) 163,
  \href{http://dx.doi.org/10.1007/JHEP01(2014)163}{\doi{10.1007/JHEP01(2014)163}},
\href{http://www.arXiv.org/abs/1311.6736}{\texttt{arXiv:1311.6736}}.

\bibitem{Khachatryan:2014doa}
\hrefCMSnoop {}{{CMS} Collaboration, ``{Search for top-squark pairs decaying
  into Higgs or Z bosons in pp collisions at $\sqrt{s}$ = 8 TeV}'',} \textit{
  Phys. Lett. B} \textbf{ 736} (2014) 371,
  \href{http://dx.doi.org/10.1016/j.physletb.2014.07.053}{\doi{10.1016/j.physletb.2014.07.053}},
\href{http://www.arXiv.org/abs/1405.3886}{\texttt{arXiv:1405.3886}}.

\bibitem{Sakai:1981gr}
\hrefCMSnoop {}{N.~Sakai, ``{Naturalness in supersymmetric GUTS}'',} \textit{
  Z. Phys. C} \textbf{ 11} (1981) 153,
\href{http://dx.doi.org/10.1007/BF01573998}{\doi{10.1007/BF01573998}}.

\bibitem{Dimopoulos:1995mi}
\hrefCMSnoop {}{S.~Dimopoulos and G.~F. Giudice, ``{Naturalness constraints in
  supersymmetric theories with nonuniversal soft terms}'',} \textit{ Phys.
  Lett. B} \textbf{ 357} (1995) 573,
  \href{http://dx.doi.org/10.1016/0370-2693(95)00961-J}{\doi{10.1016/0370-2693(95)00961-J}},
\href{http://www.arXiv.org/abs/hep-ph/9507282}{\texttt{arXiv:hep-ph/9507282}}.

\bibitem{Papucci:2011wy}
\hrefCMSnoop {}{M.~Papucci, J.~T. Ruderman, and A.~Weiler, ``{Natural SUSY
  endures}'',} \textit{ JHEP} \textbf{ 09} (2012) 035,
  \href{http://dx.doi.org/10.1007/JHEP09(2012)035}{\doi{10.1007/JHEP09(2012)035}},
\href{http://www.arXiv.org/abs/1110.6926}{\texttt{arXiv:1110.6926}}.

\bibitem{Brust:2011tb}
\hrefCMSnoop {}{C.~Brust, A.~Katz, S.~Lawrence, and R.~Sundrum, ``{SUSY, the
  Third Generation and the LHC}'',} \textit{ JHEP} \textbf{ 03} (2012) 103,
  \href{http://dx.doi.org/10.1007/JHEP03(2012)103}{\doi{10.1007/JHEP03(2012)103}},
\href{http://www.arXiv.org/abs/1110.6670}{\texttt{arXiv:1110.6670}}.

\bibitem{sms1}
\hrefCMSnoop {}{J.~Alwall, P.~Schuster, and N.~Toro, ``Simplified models for a
  first characterization of new physics at the {LHC}'',} \textit{ Phys. Rev. D}
  \textbf{ 79} (2009) 075020,
  \href{http://dx.doi.org/10.1103/PhysRevD.79.075020}{\doi{10.1103/PhysRevD.79.075020}},
\href{http://www.arXiv.org/abs/0810.3921}{\texttt{arXiv:0810.3921}}.

\bibitem{sms2}
\hrefCMSnoop {}{D.~Alves {et~al.}, ``Simplified models for {LHC} new physics
  searches'',} \textit{ J. Phys. G} \textbf{ 39} (2012) 105005,
  \href{http://dx.doi.org/10.1088/0954-3899/39/10/105005}{\doi{10.1088/0954-3899/39/10/105005}},
\href{http://www.arXiv.org/abs/1105.2838}{\texttt{arXiv:1105.2838}}.

\bibitem{CMS}
\hrefCMSnoop {}{{CMS} Collaboration, ``The {CMS} experiment at the {CERN}
  {LHC}'',} \textit{ JINST} \textbf{ 3} (2008) S08004,
\href{http://dx.doi.org/10.1088/1748-0221/3/08/S08004}{\doi{10.1088/1748-0221/3/08/S08004}}.

\bibitem{PFT-09-001}
\href {https://cdsweb.cern.ch/record/1194487}{{CMS} Collaboration, ``Particle
  Flow Event Reconstruction in CMS and Performance for Jets, Taus and MET'',}
  CMS Physics Analysis Summary CMS-PAS-PFT-09-001, 2009.

\bibitem{PFT-10-001}
\href {http://cdsweb.cern.ch/record/1247373}{{CMS} Collaboration,
  ``Commissioning of the Particle-flow Event Reconstruction with the first LHC
  collisions recorded in the CMS detector'',} CMS Physics Analysis Summary
  CMS-PAS-PFT-10-001, 2010.

\bibitem{EGM-10-004}
\href {http://cdsweb.cern.ch/record/1299116}{{CMS} Collaboration, ``{Electron
  reconstruction and identification at $\sqrt{s}$ = 7 TeV}'',} CMS Physics
  Analysis Summary CMS-PAS-EGM-10-004, 2010.

\bibitem{Chatrchyan:2013dga}
\hrefCMSnoop {}{{CMS} Collaboration, ``{Energy calibration and resolution of
  the CMS electromagnetic calorimeter in pp collisions at $\sqrt{s}=7$~TeV}'',}
  \textit{ JINST} \textbf{ 8} (2013) P09009,
  \href{http://dx.doi.org/10.1088/1748-0221/8/09/P09009}{\doi{10.1088/1748-0221/8/09/P09009}},
\href{http://www.arXiv.org/abs/1306.2016}{\texttt{arXiv:1306.2016}}.

\bibitem{CMS-PAPER-MUO-10-004}
\hrefCMSnoop {}{{CMS} Collaboration, ``Performance of {CMS} muon reconstruction
  in pp collision events at {$\sqrt{s} = 7$\TeV}'',} \textit{ J. Instrum.}
  \textbf{ 7} (2012) P10002,
  \href{http://dx.doi.org/10.1088/1748-0221/7/10/P10002}{\doi{10.1088/1748-0221/7/10/P10002}}.

\bibitem{antikt}
\hrefCMSnoop {}{M.~Cacciari, G.~P. Salam, and G.~Soyez, ``The anti-$k_t$ jet
  clustering algorithm'',} \textit{ JHEP} \textbf{ 04} (2008) 063,
  \href{http://dx.doi.org/10.1088/1126-6708/2008/04/063}{\doi{10.1088/1126-6708/2008/04/063}},
  \href{http://www.arXiv.org/abs/0802.1189}{\texttt{arXiv:0802.1189}}.

\bibitem{JETJINST}
\hrefCMSnoop {}{{CMS} Collaboration, ``{Determination of Jet Energy Calibration
  and Transverse Momentum Resolution in CMS}'',} \textit{ JINST} \textbf{ 6}
  (2011) P11002,
  \href{http://dx.doi.org/10.1088/1748-0221/6/11/P11002}{\doi{10.1088/1748-0221/6/11/P11002}},
\href{http://www.arXiv.org/abs/1107.4277}{\texttt{arXiv:1107.4277}}.

\bibitem{METPAS}
\hrefCMSnoop {}{{CMS} Collaboration, ``Missing transverse energy performance of
  the CMS detector'',} \textit{ JINST} \textbf{ 6} (2011) P09001,
  \href{http://dx.doi.org/10.1088/1748-0221/6/09/P09001}{\doi{10.1088/1748-0221/6/09/P09001}},
  \href{http://www.arXiv.org/abs/:1106.5048}{\texttt{arXiv::1106.5048}}.

\bibitem{PU_JET_AREAS}
\hrefCMSnoop {}{M.~Cacciari and G.~P. Salam, ``Pileup subtraction using jet
  areas'',} \textit{ Phys. Lett. B} \textbf{ 659} (2007) 119,
  \href{http://dx.doi.org/10.1016/j.physletb.2007.09.077}{\doi{10.1016/j.physletb.2007.09.077}},
\href{http://www.arXiv.org/abs/0707.1378}{\texttt{arXiv:0707.1378}}.

\bibitem{JET_AREAS}
\hrefCMSnoop {}{M.~Cacciari, G.~P. Salam, and G.~Soyez, ``The Catchment Area of
  Jets'',} \textit{ JHEP} \textbf{ 04} (2007) 005,
  \href{http://dx.doi.org/10.1088/1126-6708/2008/04/005}{\doi{10.1088/1126-6708/2008/04/005}},
\href{http://www.arXiv.org/abs/0802.1188}{\texttt{arXiv:0802.1188}}.

\bibitem{ref:btag}
\hrefCMSnoop {}{{CMS} Collaboration, ``{Identification of b-quark jets with the
  CMS experiment}'',} \textit{ JINST} \textbf{ 8} (2013) P04013,
  \href{http://dx.doi.org/10.1088/1748-0221/8/04/P04013}{\doi{10.1088/1748-0221/8/04/P04013}},
\href{http://www.arXiv.org/abs/1211.4462}{\texttt{arXiv:1211.4462}}.

\bibitem{ref:ref_019}
J.~Alwall\hrefCMSnoop {}{ {et~al.}, ``{MadGraph/MadEvent} v4: the new web
  generation'',} \textit{ JHEP} \textbf{ 09} (2007) 028,
  \href{http://dx.doi.org/10.1088/1126-6708/2007/09/028}{\doi{10.1088/1126-6708/2007/09/028}},
  \href{http://www.arXiv.org/abs/arXiv:0706.2334}{\texttt{arXiv:arXiv:0706.2334}}.

\bibitem{MCatNLO1}
\hrefCMSnoop {}{S.~Frixione and B.~R. Webber, ``{Matching NLO QCD computations
  and parton shower simulations}'',} \textit{ JHEP} \textbf{ 06} (2002) 029,
  \href{http://dx.doi.org/10.1088/1126-6708/2002/06/029}{\doi{10.1088/1126-6708/2002/06/029}},
\href{http://www.arXiv.org/abs/hep-ph/0204244}{\texttt{arXiv:hep-ph/0204244}}.

\bibitem{MCatNLO2}
\hrefCMSnoop {}{S.~Frixione, P.~Nason, and B.~R. Webber, ``{Matching NLO QCD
  and parton showers in heavy flavor production}'',} \textit{ JHEP} \textbf{
  08} (2003) 007,
  \href{http://dx.doi.org/10.1088/1126-6708/2003/08/007}{\doi{10.1088/1126-6708/2003/08/007}},
\href{http://www.arXiv.org/abs/hep-ph/0305252}{\texttt{arXiv:hep-ph/0305252}}.

\bibitem{Sjostrand:2006za}
\hrefCMSnoop {}{T.~Sj{\"o}strand, S.~Mrenna, and P.~Skands, ``{PYTHIA} 6.4
  physics and manual'',} \textit{ JHEP} \textbf{ 05} (2006) 026,
  \href{http://dx.doi.org/10.1088/1126-6708/2006/05/026}{\doi{10.1088/1126-6708/2006/05/026}},
\href{http://www.arXiv.org/abs/hep-ph/0603175}{\texttt{arXiv:hep-ph/0603175}}.

\bibitem{Geant}
\hrefCMSnoop {}{{GEANT4} Collaboration, ``{GEANT4}---a simulation toolkit'',}
  \textit{ Nucl. Instrum. Meth. A} \textbf{ 506} (2003) 250,
  \href{http://dx.doi.org/10.1016/S0168-9002(03)01368-8}{\doi{10.1016/S0168-9002(03)01368-8}}.

\bibitem{Abdullin:2011zz}
\hrefCMSnoop {}{{CMS} Collaboration, ``{The fast simulation of the CMS detector
  at LHC}'',} \textit{ J. Phys. Conf. Ser.} \textbf{ 331} (2011) 032049,
\href{http://dx.doi.org/10.1088/1742-6596/331/3/032049}{\doi{10.1088/1742-6596/331/3/032049}}.

\bibitem{Rahmat:2012fs}
\hrefCMSnoop {}{R.~Rahmat, R.~Kroeger, and A.~Giammanco, ``{The fast simulation
  of the CMS experiment}'',} \textit{ J. Phys. Conf. Ser.} \textbf{ 396} (2012)
  062016,
\href{http://dx.doi.org/10.1088/1742-6596/396/6/062016}{\doi{10.1088/1742-6596/396/6/062016}}.

\bibitem{ref:STAT1}
\hrefCMSnoop {}{A.~L. Read, ``Presentation of search results: the {$CL_s$}
  technique'',} \textit{ J. Phys. G} \textbf{ 28} (2002) 2693,
  \href{http://dx.doi.org/10.1088/0954-3899/28/10/313}{\doi{10.1088/0954-3899/28/10/313}}.

\bibitem{ref:STAT2}
\hrefCMSnoop {}{T.~Junk, ``{Confidence level computation for combining searches
  with small statistics}'',} \textit{ Nucl. Instrum. Meth. A} \textbf{ 434}
  (1999) 435,
  \href{http://dx.doi.org/10.1016/S0168-9002(99)00498-2}{\doi{10.1016/S0168-9002(99)00498-2}},
\href{http://www.arXiv.org/abs/hep-ex/9902006}{\texttt{arXiv:hep-ex/9902006}}.

\bibitem{LHC-HCG}
\href {http://cdsweb.cern.ch/record/1379837}{{ATLAS and CMS Collaborations, LHC
  Higgs Combination Group}, ``Procedure for the {LHC} {H}iggs boson search
  combination in {S}ummer 2011'',} Technical Report ATL-PHYS-PUB 2011-11, CMS
  NOTE 2011/005, 2011.

\bibitem{Alekhin:2011sk}
\hrefCMSnoop {}{S.~Alekhin {et~al.}, ``{The PDF4LHC Working Group Interim
  Report}'',} (2011).
\href{http://www.arXiv.org/abs/1101.0536}{\texttt{arXiv:1101.0536}}.

\bibitem{ref:PDF4LHC}
\hrefCMSnoop {}{M.~Botje {et~al.}, ``The {PDF4LHC} Working Group Interim
  Recommandations'',} (2011).
  \href{http://www.arXiv.org/abs/1101.0538}{\texttt{arXiv:1101.0538}}.

\bibitem{LUMIPAS}
\href {http://cdsweb.cern.ch/record/1598864}{{CMS} Collaboration, ``CMS
  luminosity based on pixel cluster counting --- Summer 2013 update'',} CMS
  Physics Analysis Summary CMS-PAS-LUM-13-001, 2013.

\bibitem{ref:SUSY_CROSS}
M.~Kr{\"a}mer\hrefCMSnoop {}{ {et~al.}, ``Supersymmetry production cross
  sections in pp collisions at $\sqrt{s}$ = 7 {TeV}'',} (2012).
  \href{http://www.arXiv.org/abs/1206.2892}{\texttt{arXiv:1206.2892}}.

\bibitem{ref:SUSY_CROSS2}
\hrefCMSnoop {}{W.~Beenakker, R.~H{\"o}pker, M.~Spira, and P.~M. Zerwas,
  ``Squark and gluino production at hadron colliders'',} \textit{ Nucl. Phys.
  B} \textbf{ 492} (1997) 51,
  \href{http://dx.doi.org/10.1016/S0550-3213(97)80027-2}{\doi{10.1016/S0550-3213(97)80027-2}},
\href{http://www.arXiv.org/abs/hep-ph/9610490}{\texttt{arXiv:hep-ph/9610490}}.

\bibitem{ref:SUSY_CROSS3}
\hrefCMSnoop {}{A.~Kulesza and L.~Motyka, ``Threshold Resummation for
  Squark-Antisquark and Gluino-Pair Production at the {LHC}'',} \textit{ Phys.
  Rev. Lett.} \textbf{ 102} (2009) 111802,
  \href{http://dx.doi.org/10.1103/PhysRevLett.102.111802}{\doi{10.1103/PhysRevLett.102.111802}},
\href{http://www.arXiv.org/abs/0807.2405}{\texttt{arXiv:0807.2405}}.

\bibitem{ref:SUSY_CROSS4}
\hrefCMSnoop {}{A.~Kulesza and L.~Motyka, ``{Soft gluon resummation for the
  production of gluino-gluino and squark-antisquark pairs at the {LHC}}'',}
  \textit{ Phys. Rev. D} \textbf{ 80} (2009) 095004,
  \href{http://dx.doi.org/10.1103/PhysRevD.80.095004}{\doi{10.1103/PhysRevD.80.095004}},
\href{http://www.arXiv.org/abs/0905.4749}{\texttt{arXiv:0905.4749}}.

\bibitem{ref:SUSY_CROSS5}
W.~Beenakker\hrefCMSnoop {}{ {et~al.}, ``{Soft-gluon resummation for squark and
  gluino hadroproduction}'',} \textit{ JHEP} \textbf{ 12} (2009) 041,
  \href{http://dx.doi.org/10.1088/1126-6708/2009/12/041}{\doi{10.1088/1126-6708/2009/12/041}},
\href{http://www.arXiv.org/abs/0909.4418}{\texttt{arXiv:0909.4418}}.

\end{thebibliography}\endgroup
